\documentclass[prd,aps,showpacs,epsf,floats,onecolumn]{revtex4}%
\usepackage{amssymb}
\usepackage{amsfonts}
\usepackage{amsmath}
\usepackage{graphicx}%
\providecommand{\U}[1]{\protect\rule{.1in}{.1in}}

\begin{document}
\title{\textbf{Decrease of Fisher information and the information geometry of
evolution equations for quantum mechanical probability amplitudes}}
\author{\textbf{Carlo Cafaro}$^{1}$ and \textbf{Paul M.\ Alsing}$^{2}$}
\affiliation{$^{1}$SUNY Polytechnic Institute, 12203 Albany, New York, USA}
\affiliation{$^{2}$Air Force Research Laboratory, Information Directorate, 13441 Rome, New
York, USA}

\begin{abstract}
The relevance of the concept of Fisher information is increasing in both
statistical physics and quantum computing. From a statistical mechanical
standpoint, the application of Fisher information in the kinetic theory of
gases is characterized by its decrease along the solutions of the Boltzmann
equation for Maxwellian molecules in the two-dimensional case. From a quantum
mechanical standpoint, the output state in Grover's quantum search algorithm
follows a geodesic path obtained from the Fubini-Study metric on the manifold
of Hilbert-space rays. Additionally, Grover's algorithm is specified by
constant Fisher information. In this paper, we present an information
geometric characterization of the oscillatory or monotonic behavior of
statistically parametrized squared probability amplitudes originating from
special functional forms of the Fisher information function: constant,
exponential decay, and power-law decay. Furthermore, for each case, we compute
both the computational speed and the availability loss of the corresponding
physical processes by exploiting a convenient Riemannian geometrization of
useful thermodynamical concepts. Finally, we briefly comment on the
possibility of using the proposed methods of information geometry to help
identify a suitable trade-off between speed and thermodynamic efficiency in
quantum search algorithms.

\end{abstract}

\maketitle

\bigskip\pagebreak

\section{Introduction}

The importance of the concept of Fisher information is increasing in both
classical and quantum settings, ranging from foundational aspects of
theoretical physics, including statistical physics, to quantum computing. In
Ref. \cite{frieden90}, the Fisher information was regarded as a measure of the
degree of disorder of an isolated statistical system. In particular, it was
shown that by minimizing the Fisher information subject to suitable physical
constraints, the resulting equilibrium probability density function satisfied
the correct differential equations for the system (including, among others,
Schr\"{o}dinger's wave equation, the Klein-Gordon equation, and the
Maxwell-Boltzmann law). Interestingly, in Ref. \cite{frieden90} it was
suggested that the Fisher information specifies an arrow of time that points
in the direction of decreasing accuracy for the determination of the mean
value of the statistical parameter that specifies the system. Connections
between the decrease of Fisher information and the second law of
thermodynamics were, to some extent, explored in Refs.
\cite{frieden94,frieden96}. In Ref. \cite{frieden95}, the concept of Fisher
information was employed to present a systematic approach to deriving
Lagrangians of relevance in physics. Of particular interest are the
applications of the notion of Fisher information in quantum theory. For
example, in Ref. \cite{reginatto98}, the principle of minimum Fisher
information is used to derive the many-particle time-dependent Schr\"{o}dinger
equation. In Ref. \cite{hall00}, it was proposed that the classical Fisher
information of a quantum observable is a measure of the robustness of the
observable with respect to noise. Indeed, it was shown that Fisher information
is proportional to the rate of entropy increase of the observable when the
quantum system is subjected to a Gaussian diffusive process. In Ref.
\cite{luo02}, in an effort to advance the information approach to physics by
linking the classical Lagrangian approach to mechanics and the concept of
Fisher information, a general notion of kinetic energy with respect to a
parameter was introduced and its consequences were discussed. For an extended
presentation of the role of Fisher information in physics, we refer to Ref.
\cite{frieden98}. In addition to covering foundational aspects of physics, the
use of Fisher information has also been extended to problems in statistical
physics from a more applied perspective. The application of Fisher information
to the kinetic theory of gases started with the investigation carried out by
McKean in Ref. \cite{mckean66}. In that work, the monotonic decreasing
behavior of the Fisher information was observed while studying a
one-dimensional toy-model of a Maxwellian gas. Following this line of
investigation, the decrease of Fisher information along the solutions of the
linear Fokker-Planck equation was reported by Toscani in Ref. \cite{toscani99}%
. In Ref. \cite{toscani92}, it was shown that Fisher information also
decreases along the Boltzmann equation for Maxwellian molecules in
two-dimensions. For a generalization of this finding extended to higher
dimensions, we refer to Ref. \cite{villani98a}. Finally, studying the
spatially homogeneous Landau equation for Maxwellian molecules, the
non-increasing behavior of the Fisher information was reported in Refs.
\cite{villani98b,villani98c,villani00}.

From a quantum computing viewpoint, quantum Fisher information can be
physically interpreted by observing that its square root is proportional to
the statistical speed, that is the instantaneous rate of change of the
absolute statistical distance between two pure states in the Hilbert space
(or, more generally, in the space of density operators for general mixtures)
along the path parametrized by a given statistical parameter. The absolute
statistical distance, in turn, is the maximum number of distinguishable states
along the parametrized path, optimized over all quantum measurements. The role
played by Fisher information in quantum information science is also becoming
increasingly important. First, we recall that variational principle driven
Riemannian geometrizations of Grover's original quantum search algorithms
appear in both nonadiabatically \cite{carlini06} and adiabatically
\cite{rezakhani09, rezakhani10} constrained dynamical settings. In the latter
framework, the link between the Bures metric of two density matrices
\cite{hubner92} and the Riemann metric tensor underlying the adiabatic
evolution is of particular significance. Second, we observe that it is known
that there are quantum speed limits for either isolated quantum systems
evolving (both nonadiabatically \cite{margolus98} and adiabatically
\cite{ali04}) according to a unitary dynamics or open quantum systems coupled
to an environment \cite{taddei13,adolfo13,seb13}. In the latter case, as
mentioned earlier, the Fisher information plays a key role in the geometric
interpretation of quantum speed limits of dynamical evolutions in quantum
computing based on the notion of statistical distance between quantum states,
either pure or mixed \cite{jones10, zwierz12}. In particular, when taking into
consideration open-system dynamics where dissipative effects may occur, the
temporal behavior of the Fisher information plays a key role in the
determination of a bound to the speed of evolution of the quantum system
\cite{adolfo13}. Third, we point out that dissipation may have a constructive
role in certain tasks of interest for quantum information processing
\cite{cirac09}. For example, it is known that dissipation can be used in a
constructive manner in quantum search problems
\cite{robert07,amin08,perez10,ari09}. For instance, in Ref. \cite{ari09}, it
was shown that introducing dissipation into Grover's original quantum search
algorithm has positive effects because it leads to a more robust search where
the oscillations between target and non-target items can be damped out.

The lack of a unifying theoretical framework for all the fundamental issues
outlined in the first, second, and third points motivate us to pursue here an
information geometric analysis wherein Riemannian geometry, probability
calculus and the statistical thermodynamical nature of Fisher information all
simultaneously play a crucial role. An important finding of great utility in
our proposed information geometric investigation is that the output state in
Grover's quantum search algorithm follows a geodesic path obtained from the
Fubini-Study metric on the manifold of Hilbert-space rays and additionally,
Grover's algorithm is specified by constant Fisher information
\cite{alvarez00,wadati01,cafaro12a,cafaro12b,cafaro17}.

In this paper, we use methods of information geometry to characterize the
oscillatory or monotonic behavior of statistically parametrized squared
probability amplitudes that correspond to suitably chosen functional forms of
the Fisher information function: constant, exponential decay, and power-law
decay. Moreover, for each case, we find both the computational speed and the
availability loss of the corresponding physical processes by making use of a
convenient Riemannian geometrization of useful thermodynamical concepts.
Finally, we propose the use of methods of information geometry to help
identify a suitable trade-off between speed and thermodynamic efficiency in
quantum search algorithms.

The layout of the remainder of this paper is as follows. In Sec. II, we
introduce the concept of Fisher information in both classical and quantum
information theory. In Sec. III, we use the notion of Fisher information in
order to quantify the concept of quantum distinguishability for both pure and
mixed quantum states. In Sec. IV, focusing on pure states and using
variational calculus techniques, we present an explicit derivation of the
information geometric evolution equations of quantum mechanical probability
amplitudes for arbitrary forms of the Fisher information function. In Sec. V,
we apply the main results obtained in the previous section to three special
scenarios: constant Fisher information, exponential decay and, power-law
decay. In particular, the oscillatory or monotonic behaviors of the
statistically parameterized squared probability amplitudes are reported. In
Sec. VI, we discuss the link among physical systems, Fisher information
functions, and geodesic paths on Riemannian manifolds. In Sec. VII, we first
review some basic material on a Riemannian geometric characterization of
thermodynamic concepts.\ Special attention is devoted to the concepts of
thermodynamic length and dissipated availability (or, availability loss
\cite{salamon83}) and, their link with the notion of Fisher information. Then,
for each of the three illustrative examples considered in Sec. V, we compute
both the availability loss and the computational speed of the quantum process
that corresponds to each selected functional form of the Fisher information.
Finally, our concluding remarks appear in Sec. VIII.

\section{Fisher information}

In this section, we briefly introduce the concept of Fisher information in
both classical and quantum information-theoretic settings.

\subsection{Classical framework}

In the framework of classical information theory, the Fisher
information\textbf{ }$\mathcal{F}\left(  \theta\right)  $ quantifies the
amount of information that an observable random variable\textbf{ }$X$\textbf{
}carries about an unknown parameter\textbf{ }$\theta$\textbf{ }upon which the
probability distribution\textbf{ }$p\left(  x|\theta\right)  =p_{\theta
}\left(  x\right)  $\textbf{ }depends. For a continuous random variable $X$,
the classical Fisher information $\mathcal{F}\left(  \theta\right)  $ is
defined as,%
\begin{equation}
\mathcal{F}_{\text{classical}}\left(  \theta\right)  \overset{\text{def}}%
{=}\left\langle \left(  \frac{\partial\log p\left(  x|\theta\right)
}{\partial\theta}\right)  ^{2}\right\rangle =\int p\left(  x|\theta\right)
\left(  \frac{\partial\log p\left(  x|\theta\right)  }{\partial\theta}\right)
^{2}dx\text{.} \label{djuly}%
\end{equation}
In this paper, $\log$ denotes the natural logarithmic function. We note that,
by means of simple algebra, $\mathcal{F}\left(  \theta\right)  $ in\ Eq.
(\ref{djuly}) can be rewritten in terms of the probability amplitude
$\sqrt{p\left(  x|\theta\right)  }$, a fundamental quantity in quantum theory:%
\begin{equation}
\mathcal{F}\left(  \theta\right)  =4\int\left(  \frac{\partial\sqrt{p\left(
x|\theta\right)  }}{\partial\theta}\right)  ^{2}dx\text{.}%
\end{equation}
The quantity\textbf{ }$\partial_{\theta}\left[  \log p\left(  x|\theta\right)
\right]  $\textbf{ }with\textbf{ }$\partial_{\theta}\overset{\text{def}}%
{=}\frac{\partial}{\partial\theta}$\textbf{ }in Eq.\textbf{ (}\ref{djuly}%
\textbf{) }is known as the score while the probability distribution\textbf{
}$p\left(  x|\theta\right)  $\textbf{ }is known as the likelihood function.
Observe that, exploiting the normalization condition for $p\left(
x|\theta\right)  $, the expectation value of the score is zero,%
\begin{equation}
\left\langle \partial_{\theta}\log p\left(  x|\theta\right)  \right\rangle
=0\text{.} \label{score}%
\end{equation}
Therefore, from Eqs. (\ref{djuly}) and (\ref{score}), we conclude that the
Fisher information\textbf{ }$\mathcal{F}\left(  \theta\right)  $\textbf{ }can
be regarded as the variance of the score function. For the sake of
completeness, we note that for a discrete random variable $X$, the classical
Fisher information $\mathcal{F}\left(  \theta\right)  $ is defined as,%
\begin{equation}
\mathcal{F}\left(  \theta\right)  \overset{\text{def}}{=}\sum_{i=1}^{n}%
p_{i}\left(  \frac{\partial\log p_{i}}{\partial\theta}\right)  ^{2}=\sum
_{i=1}^{n}\frac{\dot{p}_{i}^{2}}{p_{i}}\text{,} \label{cumpa1}%
\end{equation}
where $p_{i}=p_{i}\left(  x|\theta\right)  $ and $\dot{p}_{i}\overset
{\text{def}}{=}\frac{\partial p_{i}}{\partial\theta}$. In anticipation of the
formal comparison with the definition of the quantum Fisher information to be
considered in the next subsection, observe that the score function
$\partial_{\theta}\left[  \log p_{i}\left(  x|\theta\right)  \right]  $ in Eq.
(\ref{cumpa1}) satisfies the following relation,%
\begin{equation}
\frac{1}{2}\left(  p_{i}\frac{\partial\log p_{i}}{\partial\theta}%
+\frac{\partial\log p_{i}}{\partial\theta}p_{i}\right)  =\frac{\partial p_{i}%
}{\partial\theta}\text{.} \label{cumpa2}%
\end{equation}
For a detailed discussion of the intimate link between the Fisher information
and the Shannon entropy, we refer to Ref. \cite{cover06}. Finally, for an
intriguing statistical mechanical interpretation of the Fisher information, we
refer to Ref. \cite{crooks12}.

\subsection{Quantum framework}

In quantum information theory, the concept of Fisher information can be
introduced in the context of a single parameter estimation problem. This
problem concerns the inference of the value of a coupling constant $\theta$ in
the Hamiltonian $\mathcal{H}_{\theta}$,%
\begin{equation}
\mathcal{H}_{\theta}\overset{\text{def}}{=}\hslash h_{0}\theta\text{, }
\label{ham}%
\end{equation}
of a probe system by observing the evolution of the probe due to
$\mathcal{H}_{\theta}$. In Eq. (\ref{ham}), $\hslash$ is the reduced Planck
constant, $\theta$ is assumed to have units of frequency, and $h_{0}$ is a
dimensionless coupling Hamiltonian. The quantum Fisher information
$\mathcal{F}_{\text{quantum}}\left(  \theta\right)  $ is defined as
\cite{brau94},%
\begin{equation}
\mathcal{F}_{\text{quantum}}\left(  \theta\right)  \overset{\text{def}}{=}%
\max_{\left\{  \mathcal{X}\left(  x\right)  \right\}  }\left[  \mathcal{F}%
\left(  \theta\right)  \right]  \text{,} \label{giannino}%
\end{equation}
with $\mathcal{F}\left(  \theta\right)  $ given by,%
\begin{equation}
\mathcal{F}\left(  \theta\right)  \overset{\text{def}}{=}\int p\left(
x|\theta\right)  \left(  \frac{\partial\log p\left(  x|\theta\right)
}{\partial\theta}\right)  ^{2}dx\text{.} \label{feta}%
\end{equation}
The quantity $\left\{  \mathcal{X}\left(  x\right)  \right\}  $ in Eq.
(\ref{giannino}) denotes a generalized measurement where $\mathcal{X}\left(
x\right)  $ are non-negative, Hermitian operators that satisfy the
completeness relation,%
\begin{equation}
\int\mathcal{X}\left(  x\right)  dx=\mathbf{1}\text{,}%
\end{equation}
with $\mathbf{1}$ denoting the unit operator. Furthermore, the probability
distribution $p\left(  x|\theta\right)  $ in Eq. (\ref{feta}) is defined as,%
\begin{equation}
p\left(  x|\theta\right)  \overset{\text{def}}{=}\text{\textrm{tr}}\left[
\mathcal{X}\left(  x\right)  \rho\left(  \theta\right)  \right]  \text{,}
\label{piro}%
\end{equation}
where $x$ labels the outcomes of the measurement and it need not be a single
continuous real variable. It can also be discrete or multivariate, for
instance. The symbol \textquotedblleft\textrm{tr}\textquotedblright\ in Eq.
(\ref{piro}) denotes the usual trace operation. The quantity $\rho\left(
\theta\right)  $ in Eq. (\ref{piro}) denotes a curve on the space of density
operators parametrized by the parameter $\theta$. Observe that while the
classical distinguishability metric satisfies the relation,%
\begin{equation}
ds_{\text{PD}}^{2}=\mathcal{F}_{\text{classical}}\left(  \theta\right)
d\theta^{2}\text{,} \label{cmetric}%
\end{equation}
the quantum distinguishability metric fulfills the condition,%
\begin{equation}
ds_{\text{DO}}^{2}=\mathcal{F}_{\text{quantum}}\left(  \theta\right)
d\theta^{2}\text{.} \label{qmetric}%
\end{equation}
Note that PD\ in Eq. (\ref{cmetric}) and DO in Eq.(\ref{qmetric}) denote
probability distributions and density operators, respectively. Braunstein and
Caves showed that $\mathcal{F}_{\text{quantum}}\left(  \theta\right)  $ can be
written as \cite{brau94},
\begin{equation}
\mathcal{F}_{\text{quantum}}\left(  \theta\right)  =\left\langle L^{2}\left(
\theta\right)  \right\rangle \overset{\text{def}}{=}\text{\textrm{tr}}\left[
\rho\left(  \theta\right)  L^{2}\left(  \theta\right)  \right]  \text{,}
\label{cumpa11}%
\end{equation}
where $L$ is the so-called symmetric logarithmic derivative operator. This
operator is defined implicitly in terms of the following relation,%
\begin{equation}
\frac{1}{2}\left(  \rho L+L\rho\right)  =\frac{\partial\rho}{\partial\theta
}\text{,} \label{cumpa22}%
\end{equation}
with,%
\begin{equation}
\frac{\partial\rho}{\partial\theta}=-i\left[  T\left(  \theta\right)  \text{,
}\rho\left(  \theta\right)  \right]  \text{,} \label{tete}%
\end{equation}
where $i$ is the imaginary unit. By replacing both the trace with the integral
(or, summation) and the density operator with the probability density
function, we observe the formal analogies between Eqs. (\ref{cumpa1}) and
(\ref{cumpa11}), and Eqs. (\ref{cumpa2}) and (\ref{cumpa22}), respectively.
The quantity $T\left(  \theta\right)  =T_{\theta}$ in Eq. (\ref{tete}) is the
Hermitian generator of displacements in the parameter $\theta$ defined as,%
\begin{equation}
T_{\theta}\left(  t\right)  \overset{\text{def}}{=}i\frac{\partial U_{\theta
}\left(  t\right)  }{\partial\theta}U_{\theta}^{\dagger}\left(  t\right)
\text{.} \label{toti}%
\end{equation}
The unitary evolution operator $U_{\theta}\left(  t\right)  $ is generated by
the Hamiltonian $\mathcal{H}_{\theta}\left(  t\right)  $,%
\begin{equation}
\mathcal{H}_{\theta}\left(  t\right)  U_{\theta}\left(  t\right)
=i\hslash\frac{\partial U_{\theta}\left(  t\right)  }{\partial t}\text{,}%
\end{equation}
where,%
\begin{equation}
\rho_{\theta}\left(  0\right)  \rightarrow\rho_{\theta}\left(  t\right)
\overset{\text{def}}{=}U_{\theta}\left(  t\right)  \rho_{\theta}\left(
0\right)  U_{\theta}^{\dagger}\left(  t\right)  \text{.}%
\end{equation}
The dagger symbol \textquotedblleft$\dagger$\textquotedblright\ in Eq.
(\ref{toti}) denotes the usual Hermitian conjugate operation. Observe that if
$\mathcal{H}_{\theta}\left(  t\right)  =\hslash h_{0}\theta$ is a constant
quantity, using Eq. (\ref{toti}), one finds that $T_{\theta}\left(  t\right)
=h_{0}t$. Then, for pure states $\rho_{\theta}^{2}=\rho_{\theta}$, it can be
shown that \cite{boi07},%
\begin{equation}
\mathcal{F}_{\text{quantum}}\left(  \theta\right)  =4\sigma_{T_{\theta}\left(
t\right)  }^{2}=4\left(  \left\langle T_{\theta}^{2}\left(  t\right)
\right\rangle -\left\langle T_{\theta}\left(  t\right)  \right\rangle
^{2}\right)  \text{.} \label{becomes}%
\end{equation}
For the sake of completeness, we remark that in the case of mixed states, the
variance provides an upper bound on the quantum Fisher information
\cite{brau96}. Furthermore, in the case of time-estimation, we have%
\begin{equation}
\theta\mapsto t\text{, }T_{\theta}\left(  t\right)  \mapsto\mathcal{H}\left(
t\right)  \text{, }\mathcal{F}_{\text{quantum}}\left(  \theta\right)
\mapsto\mathcal{F}_{\text{quantum}}\left(  t\right)  \text{,}%
\end{equation}
and Eq. (\ref{becomes}) becomes%
\begin{equation}
\mathcal{F}_{\text{quantum}}\left(  t\right)  =\frac{4}{\hslash^{2}}%
\sigma_{\mathcal{H}\left(  t\right)  }^{2}=\frac{4}{\hslash^{2}}\left(
\left\langle \mathcal{H}^{2}\left(  t\right)  \right\rangle -\left\langle
\mathcal{H}\left(  t\right)  \right\rangle ^{2}\right)  \text{.}%
\end{equation}
The quantum Fisher information $\mathcal{F}_{\text{quantum}}\left(
\theta\right)  $ can be interpreted in an efficient manner as the square of a
statistical speed $v_{\mathcal{F}}$ \cite{taddei13,pezze09}:%
\begin{equation}
\mathcal{F}_{\text{quantum}}\left(  \theta\right)  =v_{\mathcal{F}}%
^{2}\overset{\text{def}}{=}\left(  \frac{dl\left(  \theta\right)  }{d\theta
}\right)  ^{2}\text{.} \label{vafa}%
\end{equation}
The quantity $v_{\mathcal{F}}$ in Eq. (\ref{vafa}) denotes the rate of change
with respect to the parameter $\theta$ of the absolute statistical distance
$l\left(  \theta\right)  $ between two pure states (or, in general, density
operators for general mixtures) in the Hilbert space. The absolute statistical
distance $l\left(  \theta\right)  $ equals the maximum number of
distinguishable states along the path $\rho\left(  \theta\right)
=\rho_{\theta}$ parametrized by $\theta$, optimized over all possible
generalized quantum measurements. For further details on the quantum Fisher
information, we refer to Refs. \cite{brau96,luo03,durkin07,boixo08}.

\section{Information geometry and quantum distinguishability}

In this section, we briefly present suitable information geometric measures of
quantum distinguishability for both pure and mixed states.

\subsection{Pure states}

Classical probability distributions can be distinguished by means of the
so-called classical Fisher-Rao information metric tensor $g_{ij}^{\left(
\text{FR}\right)  }\left(  \theta\right)  $ given by \cite{amari},%
\begin{equation}
g_{ij}^{\left(  \text{FR}\right)  }\left(  \theta\right)  \overset{\text{def}%
}{=}\int p\left(  x|\theta\right)  \frac{\partial\log\left[  p\left(
x|\theta\right)  \right]  }{\partial\theta^{i}}\frac{\partial\log\left[
p\left(  x|\theta\right)  \right]  }{\partial\theta^{j}}dx=4\int\frac
{\partial\sqrt{p\left(  x|\theta\right)  }}{\partial\theta^{i}}\frac
{\partial\sqrt{p\left(  x|\theta\right)  }}{\partial\theta^{j}}dx\text{.}
\label{FR1}%
\end{equation}
One possible way of transitioning from the classical to the quantum settings
is to replace the integral and the probability density function $p\left(
x|\theta\right)  =p_{\theta}\left(  x\right)  =p_{\theta}$ in Eq. (\ref{FR1})
with the trace operation and the density operator $\rho_{\theta}$,
respectively. Then, the quantum version of $g_{ij}^{\left(  \text{FR}\right)
}\left(  \theta\right)  $ in Eq. (\ref{FR1}) becomes the so-called
Wigner-Yanase metric $g_{ij}^{\left(  \text{WY}\right)  }\left(
\theta\right)  $ \cite{luo03,luo06},
\begin{equation}
g_{ij}^{\left(  \text{WY}\right)  }\left(  \theta\right)  =4\text{\textrm{tr}%
}\left[  \left(  \partial_{i}\sqrt{\rho_{\theta}}\right)  \left(  \partial
_{j}\sqrt{\rho_{\theta}}\right)  \right]  =4\text{\textrm{tr}}\left[  \left(
\partial_{i}\rho_{\theta}\right)  \left(  \partial_{j}\rho_{\theta}\right)
\right]  \text{,} \label{WY}%
\end{equation}
since $\rho_{\theta}=\rho_{\theta}^{2}$ with $\rho_{\theta}$ being\textbf{ }a
pure state. As pointed out in Ref. \cite{luo03}, quantum generalizations of
the Fisher information are not unique. Observe that $\partial_{i}\rho_{\theta
}$ in Eq. (\ref{WY}) can be written as,%
\begin{equation}
\partial_{i}\rho_{\theta}=\partial_{i}\left(  \left\vert \psi_{\theta
}\right\rangle \left\langle \psi_{\theta}\right\vert \right)  =\left\vert
\partial_{i}\psi_{\theta}\right\rangle \left\langle \psi_{\theta}\right\vert
+\left\vert \psi_{\theta}\right\rangle \left\langle \partial_{i}\psi_{\theta
}\right\vert \text{.}%
\end{equation}
Therefore, after some straightforward algebra, we find
\begin{align}
\left(  \partial_{i}\rho_{\theta}\right)  \left(  \partial_{j}\rho_{\theta
}\right)   &  =\left\langle \psi_{\theta}|\partial_{j}\psi_{\theta
}\right\rangle \left\vert \partial_{i}\psi_{\theta}\right\rangle \left\langle
\psi_{\theta}\right\vert +\left\vert \partial_{i}\psi_{\theta}\right\rangle
\left\langle \partial_{j}\psi_{\theta}\right\vert +\left\langle \partial
_{i}\psi_{\theta}|\partial_{j}\psi_{\theta}\right\rangle \left\vert
\psi_{\theta}\right\rangle \left\langle \psi_{\theta}\right\vert +\nonumber\\
& \nonumber\\
&  +\left\langle \partial_{i}\psi_{\theta}|\psi_{\theta}\right\rangle
\left\vert \psi_{\theta}\right\rangle \left\langle \partial_{j}\psi_{\theta
}\right\vert \text{.} \label{doppio}%
\end{align}
Using Eq. (\ref{doppio}), \textrm{tr}$\left[  \left(  \partial_{i}\rho
_{\theta}\right)  \left(  \partial_{j}\rho_{\theta}\right)  \right]  $ in Eq.
(\ref{WY}) can be recast as%
\begin{align}
\text{\textrm{tr}}\left[  \left(  \partial_{i}\rho_{\theta}\right)  \left(
\partial_{j}\rho_{\theta}\right)  \right]   &  =\left\langle \psi_{\theta
}|\left(  \partial_{i}\rho_{\theta}\right)  \left(  \partial_{j}\rho_{\theta
}\right)  |\psi_{\theta}\right\rangle \nonumber\\
& \nonumber\\
&  =\left\langle \psi_{\theta}|\partial_{j}\psi_{\theta}\right\rangle
\left\langle \psi_{\theta}|\partial_{i}\psi_{\theta}\right\rangle
+\left\langle \psi_{\theta}|\partial_{i}\psi_{\theta}\right\rangle
\left\langle \partial_{j}\psi_{\theta}|\psi_{\theta}\right\rangle +\nonumber\\
& \nonumber\\
&  +\left\langle \partial_{i}\psi_{\theta}|\partial_{j}\psi_{\theta
}\right\rangle +\left\langle \partial_{i}\psi_{\theta}|\psi_{\theta
}\right\rangle \left\langle \partial_{j}\psi_{\theta}|\psi_{\theta
}\right\rangle \text{.} \label{quisopra}%
\end{align}
Using the normalization condition $\left\langle \psi_{\theta}|\psi_{\theta
}\right\rangle =1$, we have $\left\langle \partial_{j}\psi_{\theta}%
|\psi_{\theta}\right\rangle =-\left\langle \psi_{\theta}|\partial_{j}%
\psi_{\theta}\right\rangle $. Therefore, \textrm{tr}$\left[  \left(
\partial_{i}\rho_{\theta}\right)  \left(  \partial_{j}\rho_{\theta}\right)
\right]  $ in Eq. (\ref{quisopra}) becomes,%
\begin{equation}
\text{\textrm{tr}}\left[  \left(  \partial_{i}\rho_{\theta}\right)  \left(
\partial_{j}\rho_{\theta}\right)  \right]  =\left\langle \partial_{i}%
\psi_{\theta}|\partial_{j}\psi_{\theta}\right\rangle +\left\langle
\partial_{i}\psi_{\theta}|\psi_{\theta}\right\rangle \left\langle \partial
_{j}\psi_{\theta}|\psi_{\theta}\right\rangle \text{.} \label{1}%
\end{equation}
Following the line of reasoning presented in Ref. \cite{provost80}, we observe
that we can write the inner product $\left\langle \partial_{i}\psi_{\theta
}|\partial_{j}\psi_{\theta}\right\rangle $ as,
\begin{equation}
\left\langle \partial_{i}\psi_{\theta}|\partial_{j}\psi_{\theta}\right\rangle
=\gamma_{ij}+i\sigma_{ij}\text{,} \label{2}%
\end{equation}
where,%
\begin{equation}
\gamma_{ij}\overset{\text{def}}{=}\operatorname{Re}\left[  \left\langle
\partial_{i}\psi_{\theta}|\partial_{j}\psi_{\theta}\right\rangle \right]
\text{, and }\sigma_{ij}\overset{\text{def}}{=}\operatorname{Im}\left[
\left\langle \partial_{i}\psi_{\theta}|\partial_{j}\psi_{\theta}\right\rangle
\right]  \text{, } \label{3}%
\end{equation}
respectively. Note that $\operatorname{Re}\left(  z\right)  $ and
$\operatorname{Im}\left(  z\right)  $ denote the real and the imaginary part
of a complex quantity $z$, respectively. Observe that $\gamma_{ij}$ and
$\sigma_{ij}$ are symmetric and antisymmetric quantities, respectively.
Indeed,
\begin{equation}
\gamma_{ji}=\operatorname{Re}\left[  \left\langle \partial_{j}\psi_{\theta
}|\partial_{i}\psi_{\theta}\right\rangle \right]  =\operatorname{Re}\left[
\left\langle \partial_{i}\psi_{\theta}|\partial_{j}\psi_{\theta}\right\rangle
^{\ast}\right]  =\operatorname{Re}\left[  \left\langle \partial_{i}%
\psi_{\theta}|\partial_{j}\psi_{\theta}\right\rangle \right]  =\gamma
_{ij}\text{,}%
\end{equation}
and,%
\begin{equation}
\sigma_{ji}=\operatorname{Im}\left[  \left\langle \partial_{j}\psi_{\theta
}|\partial_{i}\psi_{\theta}\right\rangle \right]  =\operatorname{Im}\left[
\left\langle \partial_{i}\psi_{\theta}|\partial_{j}\psi_{\theta}\right\rangle
^{\ast}\right]  =-\operatorname{Im}\left[  \left\langle \partial_{i}%
\psi_{\theta}|\partial_{j}\psi_{\theta}\right\rangle \right]  =-\sigma
_{ij}\text{.}%
\end{equation}
Since $\sigma_{ij}=-\sigma_{ji}$, $\sigma_{ij}d\theta^{i}d\theta^{j}=0$.
Finally, by\textbf{ }using Eqs. (\ref{1}), (\ref{2}) and (\ref{3}),
$g_{ij}^{\left(  \text{WY}\right)  }\left(  \theta\right)  $ in Eq. (\ref{WY})
becomes
\begin{equation}
g_{ij}^{\left(  \text{WY}\right)  }\left(  \theta\right)  =4\left\{
\operatorname{Re}\left[  \left\langle \partial_{i}\psi_{\theta}|\partial
_{j}\psi_{\theta}\right\rangle \right]  +\left\langle \partial_{i}\psi
_{\theta}|\psi_{\theta}\right\rangle \left\langle \partial_{j}\psi_{\theta
}|\psi_{\theta}\right\rangle \right\}  \text{.}%
\end{equation}
For the sake of completeness, we recall that%
\begin{equation}
g_{ij}^{\left(  \text{FS}\right)  }\left(  \theta\right)  =\frac{1}{4}%
g_{ij}^{\left(  \text{WY}\right)  }\left(  \theta\right)  \text{,}
\label{fswy}%
\end{equation}
where $g_{ij}^{\left(  \text{FS}\right)  }\left(  \theta\right)  $ denotes the
Fubini-Study metric. The infinitesimal line element $ds_{\text{FS}}^{2}$
corresponding to the Fubini-Study metric tensor $g_{ij}^{\left(
\text{FS}\right)  }\left(  \theta\right)  $ is given by,%
\begin{equation}
ds_{\text{FS}}^{2}=g_{ij}^{\left(  \text{FS}\right)  }\left(  \theta\right)
d\theta^{i}d\theta^{j}\text{.}%
\end{equation}
The metric tensor components $g_{ij}^{\left(  \text{FS}\right)  }\left(
\theta\right)  $ must be such that \cite{provost80}: (1) they transform
properly under a change of the coordinates $\theta\rightarrow\theta^{\prime
}=\theta^{\prime}\left(  \theta\right)  $, (2) they are invariant under gauge
transformations, $\psi\left(  \theta\right)  \rightarrow\psi^{\prime}\left(
\theta\right)  =e^{i\alpha\left(  \theta\right)  }\psi\left(  \theta\right)
$, and (3) they define a positive definite metric tensor. Imposing these
conditions, it can be shown that $ds_{\text{FS}}^{2}$ can be defined as,%
\begin{equation}
ds_{\text{FS}}^{2}\overset{\text{def}}{=}\left\Vert d\psi\right\Vert
^{2}-\left\vert \left\langle \psi|d\psi\right\rangle \right\vert
^{2}=\left\langle d\psi|d\psi\right\rangle -\left\langle d\psi|\psi
\right\rangle \left\langle \psi|d\psi\right\rangle =\left\langle d\psi_{\bot
}|d\psi_{\bot}\right\rangle =1-|\left\langle \psi^{\prime}|\psi\right\rangle
|^{2}\text{,} \label{fsmetric}%
\end{equation}
where $\left\vert d\psi\right\rangle $ and $\left\vert d\psi_{\bot
}\right\rangle $ are given by,%
\begin{equation}
\left\vert d\psi\right\rangle \overset{\text{def}}{=}\left\vert \psi^{\prime
}\right\rangle -\left\vert \psi\right\rangle \text{, and }\left\vert
d\psi_{\bot}\right\rangle \overset{\text{def}}{=}\left\vert d\psi\right\rangle
-\left\vert \psi\right\rangle \left\langle \psi|d\psi\right\rangle \text{,}
\label{defi}%
\end{equation}
respectively. For the sake of clarity, we remark that $\left\vert
\psi\right\rangle $ and $\left\vert \psi^{\prime}\right\rangle $ are two
neighboring normalized pure states, $\left\vert d\psi\right\rangle $ is the
difference between them, and $\left\vert d\psi_{\bot}\right\rangle $ is the
projection of $\left\vert d\psi\right\rangle $ orthogonal to $\left\vert
\psi\right\rangle $. Expanding $\left\vert \psi\right\rangle $ and $\left\vert
\psi^{\prime}\right\rangle $ with respect to an orthonormal basis $\left\{
\left\vert m\right\rangle \right\}  $ with $m\in\left\{  1\text{,...,
}N\right\}  $, we obtain%
\begin{equation}
\left\vert \psi\right\rangle \overset{\text{def}}{=}\sum_{m=1}^{N}\sqrt
{p_{m}\left(  \theta\right)  }e^{i\phi_{m}\left(  \theta\right)  }\left\vert
m\right\rangle \text{ and }\left\vert \psi^{\prime}\right\rangle
\overset{\text{def}}{=}\sum_{m=1}^{N}\sqrt{p_{m}+dp_{m}}e^{i\left(  \phi
_{m}+d\phi_{m}\right)  }\left\vert m\right\rangle \text{,} \label{explicit}%
\end{equation}
respectively. Substituting Eq. (\ref{explicit}) into Eq. (\ref{fsmetric}) and
recalling Eq. (\ref{fswy}), after some tedious but straightforward algebra
\cite{cafaro12b}, the infinitesimal Wigner-Yanase line element $ds_{\text{WY}%
}^{2}=4ds_{\text{FS}}^{2}$ becomes,
\begin{equation}
ds_{\text{WY}}^{2}=\left\{  \sum_{m=1}^{N}\frac{\dot{p}_{m}^{2}}{p_{m}%
}+4\left[  \sum_{m=1}^{N}p_{m}\dot{\phi}_{m}^{2}-\left(  \sum_{m=1}^{N}%
p_{m}\dot{\phi}_{m}\right)  ^{2}\right]  \right\}  d\theta^{2}\text{,}
\label{dwy}%
\end{equation}
where,%
\begin{equation}
\dot{p}_{m}\overset{\text{def}}{=}\frac{dp_{m}}{d\theta}\text{ and, }\dot
{\phi}_{m}\overset{\text{def}}{=}\frac{d\phi_{m}}{d\theta}\text{.}
\label{okoggi}%
\end{equation}
In the next subsection, we move our discussion from pure states to density operators.

\subsection{Density operators}

In the case of density operators, one needs to consider the quantum analog
$\mathcal{M}_{\vec{\rho}}$ of the probability simplex \cite{sam,cafaro12},
\begin{equation}
\mathcal{M}_{\vec{\rho}}\overset{\text{def}}{=}\left\{  \vec{\rho}%
\in\mathcal{L}\left(  \mathcal{H}\right)  :\vec{\rho}\overset{\text{def}}%
{=}\sum_{i\text{, }j=1}^{N}\rho^{ij}\vec{e}_{ij}\text{, }\vec{\rho}=\vec{\rho
}^{\dagger}\text{, \textrm{tr}}\left(  \vec{\rho}\right)  =1\text{, }\vec
{\rho}\text{ }\geq0\right\}  \text{,} \label{mro}%
\end{equation}
where $\mathcal{L}\left(  \mathcal{H}\right)  $ denotes the linear space of
all linear operators on a $N$-dimensional Hilbert space $\mathcal{H}$ with
density operators $\vec{\rho}$ written as vectors in $\mathcal{L}\left(
\mathcal{H}\right)  $. The space $\mathcal{M}_{\vec{\rho}}$ in Eq. (\ref{mro})
is an $\left(  N^{2}-1\right)  $-dimensional \emph{real} manifold with
nontrivial boundary. An arbitrary linear operator vector $\vec{V}$ on
$\mathcal{H}$ can be decomposed with respect to an operator vector basis
$\vec{e}_{ij}\overset{\text{def}}{=}\left\vert i\right\rangle \left\langle
j\right\vert $ with $i$, $j=1$,..., $N$ as,%
\begin{equation}
\vec{V}=\sum_{i\text{, }j=1}^{N}\left\langle i|\vec{V}|j\right\rangle \vec
{e}_{ij}\text{ }=\sum_{i\text{, }j=1}^{N}V^{ij}\vec{e}_{ij}\text{ .}%
\end{equation}
The tangent space at $\vec{\rho}$ is characterized by an $\left(
N^{2}-1\right)  $-dimensional \emph{real} vector space of traceless Hermitian
operators $\vec{T}$,%
\begin{equation}
\vec{T}=\sum_{i\text{, }j=1}^{N}T^{ij}\vec{e}_{ij}\text{, }%
\end{equation}
with \textrm{tr}$\left(  \vec{T}\right)  =0$. The action of $1$-forms
$\tilde{F}$, expanded in terms of the dual basis $\tilde{\omega}^{ji}%
\overset{\text{def}}{=}\left\vert i\right\rangle \left\langle j\right\vert $,%
\begin{equation}
\tilde{F}\overset{\text{def}}{=}\sum_{i\text{, }j=1}^{N}F_{ij}\tilde{\omega
}^{ji}\text{,}%
\end{equation}
on density operators $\vec{\rho}$ is defined as,%
\begin{equation}
\tilde{F}\left(  \vec{\rho}\right)  \overset{\text{def}}{=}\left\langle
\tilde{F}\text{, }\vec{\rho}\right\rangle =\sum_{i\text{, }j\text{, }l\text{,
}k=1}^{N}F_{ij}\rho^{lk}\left\langle \tilde{\omega}^{ji}\text{, }\vec{e}%
_{lk}\right\rangle =\sum_{i\text{, }j\text{, }l\text{, }k=1}^{N}F_{ij}%
\rho^{lk}\delta_{l}^{j}\delta_{k}^{i}=\sum_{i\text{, }j=1}^{N}F_{ij}\rho
^{ji}=\text{\textrm{tr}}\left(  \tilde{F}\vec{\rho}\right)  \overset
{\text{def}}{=}\left\langle \tilde{F}\right\rangle \text{.} \label{observable}%
\end{equation}
Therefore, from Eq. (\ref{observable}), a Hermitian $1$-form $\tilde{F}%
=\tilde{F}^{\dagger}$ is an ordinary quantum observable with $\left\langle
\tilde{F}\text{, }\vec{\rho}\right\rangle =\left\langle \tilde{F}\right\rangle
$. A metric structure $\mathbf{g}_{\vec{\rho}}\left(  \cdot\text{, }%
\cdot\right)  $ on the manifold $\mathcal{M}_{\vec{\rho}}$ can be introduced
by specifying the action of the metric on a pair of $1$-forms $\tilde{A}$ and
$\tilde{B}$ as,%
\begin{equation}
\mathbf{g}_{\vec{\rho}}\left(  \tilde{A}\text{, }\tilde{B}\right)
\overset{\text{def}}{=}\left\langle \frac{\tilde{A}\tilde{B}+\tilde{B}%
\tilde{A}}{2}\right\rangle =\text{\textrm{tr}}\left[  \left(  \frac{\tilde
{A}\tilde{B}+\tilde{B}\tilde{A}}{2}\right)  \vec{\rho}\right]
=\text{\textrm{tr}}\left[  \frac{\tilde{A}}{2}\left(  \vec{\rho}\tilde
{B}+\tilde{B}\vec{\rho}\right)  \right]  =\left\langle \tilde{A}\text{,
}\mathcal{R}_{\vec{\rho}}\left(  \tilde{B}\right)  \right\rangle \text{,}
\label{quantumm}%
\end{equation}
where $\mathcal{R}_{\vec{\rho}}\left(  \tilde{B}\right)  $ is the raising
operator that maps $1$-forms (lower covariant components) to vectors (upper
contravariant components) \cite{sam},
\begin{equation}
\mathcal{R}_{\vec{\rho}}\left(  \tilde{B}\right)  \overset{\text{def}}{=}%
\frac{\vec{\rho}\tilde{B}+\tilde{B}\vec{\rho}}{2}\text{.}%
\end{equation}
The metric $\mathbf{g}_{\vec{\rho}}\left(  \tilde{A}\text{, }\tilde{B}\right)
$ in Eq. (\ref{quantumm}) is constructed in terms of statistical correlations
of quantum observables. By means of the lowering operator $\mathcal{L}%
_{\vec{\rho}}\left(  \vec{A}\right)  $ that maps vectors to $1$-forms
\cite{sam},%
\begin{equation}
\mathcal{L}_{\vec{\rho}}\left(  \vec{A}\right)  \overset{\text{def}}%
{=}\mathcal{R}_{\vec{\rho}}^{-1}\left(  \vec{A}\right)  \text{,}%
\end{equation}
the action of the metric tensor $g_{\vec{\rho}}\left(  \cdot\text{, }%
\cdot\right)  $ on a pair of vectors $\vec{A}$ and $\vec{B}$ can be defined
as,%
\begin{equation}
\mathbf{g}_{\vec{\rho}}\left(  \vec{A}\text{, }\vec{B}\right)  \overset
{\text{def}}{=}\left\langle \mathcal{L}_{\vec{\rho}}\left(  \vec{A}\right)
\text{, }\vec{B}\right\rangle =\text{\textrm{tr}}\left[  \vec{B}%
\mathcal{L}_{\vec{\rho}}\left(  \vec{A}\right)  \right]  \text{.}%
\end{equation}
Finally, the quantum line element for density operators is given by,%
\begin{equation}
ds_{\text{DO}}^{2}\overset{\text{def}}{=}\mathbf{g}_{\vec{\rho}}\left(
d\vec{\rho}\text{, }d\vec{\rho}\right)  \text{,} \label{qm}%
\end{equation}
where $d\vec{\rho}\overset{\text{def}}{=}\left(  \vec{\rho}+d\vec{\rho
}\right)  -\vec{\rho}$ with,%
\begin{equation}
\vec{\rho}\overset{\text{def}}{=}\sum_{j=1}^{N}p^{_{j}}\left\vert
j\right\rangle \left\langle j\right\vert \text{, and }\vec{\rho}+d\vec{\rho
}\overset{\text{def}}{=}\sum_{j=1}^{N}\left(  p^{_{j}}+dp^{_{j}}\right)
\left\vert j^{\prime}\right\rangle \left\langle j^{\prime}\right\vert \text{,}%
\end{equation}
and, $\left\vert j^{\prime}\right\rangle \overset{\text{def}}{=}e^{id\theta
h}\left\vert j\right\rangle $. The quantity $e^{id\theta h}$ denotes an
infinitesimal unitary transformation on the orthonormal basis that
diagonalizes $\vec{\rho}$ while $h$ is the Hermitian operator that generates
the infinitesimal unitary basis transformations. After some algebra,
$d\vec{\rho}$ can be rewritten as%
\begin{equation}
d\vec{\rho}\overset{\text{def}}{=}\sum_{j=1}^{N}dp^{j}\left\vert
j\right\rangle \left\langle j\right\vert +id\theta\sum_{j\text{,}\ k=1}%
^{N}\left(  p^{j}-p^{k}\right)  h_{kj}\left\vert k\right\rangle \left\langle
j\right\vert \text{,} \label{diro}%
\end{equation}
where $h_{kj}\overset{\text{def}}{=}\left\langle k|h|j\right\rangle $.
Finally, by\textbf{ }substituting Eq. (\ref{diro}) into Eq. (\ref{qm}),
$ds_{\text{DO}}^{2}$ becomes \cite{sam}%
\begin{equation}
ds_{\text{DO}}^{2}=\mathbf{g}_{\vec{\rho}}\left(  d\vec{\rho}\text{, }%
d\vec{\rho}\right)  \overset{\text{def}}{=}\text{\textrm{tr}}\left[
d\vec{\rho}\mathcal{L}_{\vec{\rho}}\left(  d\vec{\rho}\right)  \right]
=\left[  \sum_{k=1}^{N}\frac{1}{p^{k}}\left(  \frac{dp^{k}}{d\theta}\right)
^{2}+2\sum_{j\neq k}\frac{\left(  p^{j}-p^{k}\right)  ^{2}}{\left(
p^{j}+p^{k}\right)  }\left\vert h_{jk}\right\vert ^{2}\right]  d\theta
^{2}\text{.} \label{qle}%
\end{equation}
Notice that the quantum line element in Eq. (\ref{qle}) is identical to the
distinguishability metric for density operators obtained by Braunstein and
Caves in Ref. \cite{brau94} by optimizing over all generalized quantum
measurements for distinguishing among neighboring quantum states. We also
point out that for pure states, the line element in Eq. (\ref{qle}) becomes
the usual Fubini-Study metric, a gauge invariant metric on the \emph{complex}
projective Hilbert space \cite{provost80}. For the sake of completeness, we
also point out that $ds_{\text{DO}}^{2}$ in Eq. (\ref{qle}) was originally
regarded as the infinitesimal form of a distance between density operators in
Ref. \cite{bures} and interpreted as a generalization of transition
probabilities to mixed states in Ref. \cite{uhlmann}. Indeed, the Bures
distance $d_{\text{Bures}}\left(  \rho_{1}\text{, }\rho_{2}\right)  $ between
two mixed density operators $\rho_{1}$ and $\rho_{2}$ is given by
\cite{bures},%
\begin{equation}
d_{\text{Bures}}\left(  \rho_{1}\text{, }\rho_{2}\right)  \overset{\text{def}%
}{=}\sqrt{2}\left[  1-\mathcal{F}\left(  \rho_{1}\text{, }\rho_{2}\right)
\right]  ^{\frac{1}{2}}\text{,}%
\end{equation}
where $\mathcal{F}\left(  \rho_{1}\text{, }\rho_{2}\right)  $ is the so-called
Uhlmann fidelity defined as \cite{uhlmann},%
\begin{equation}
\mathcal{F}\left(  \rho_{1}\text{, }\rho_{2}\right)  \overset{\text{def}}%
{=}\mathrm{tr}\sqrt{\sqrt{\rho_{1}}\rho_{2}\sqrt{\rho_{1}}}\text{.}%
\end{equation}
The infinitesimal Bures line element $ds_{\text{Bures}}^{2}$ can be expressed
in terms of the Bures distance between two infinitesimally close density
matrices as \cite{hubner92},%
\begin{equation}
ds_{\text{Bures}}^{2}\overset{\text{def}}{=}d_{\text{Bures}}^{2}\left(
\rho\text{, }\rho+d\rho\right)  =\frac{1}{2}\sum_{i,j}\frac{\left\vert
\left\langle i|d\rho|j\right\rangle \right\vert ^{2}}{p^{i}+p^{j}}\text{,}
\label{dbures}%
\end{equation}
where $\rho$ and $d\rho$ in Eq. (\ref{dbures}) are defined as,%
\begin{equation}
\rho\overset{\text{def}}{=}\sum_{i=1}^{N}p^{i}\left\vert i\right\rangle
\left\langle i\right\vert \text{ and, }d\rho\overset{\text{def}}{=}\sum
_{i=1}^{N}dp^{i}\left\vert i\right\rangle \left\langle i\right\vert
+\sum_{i=1}^{N}p^{i}\left\vert di\right\rangle \left\langle i\right\vert
+\sum_{i=1}^{N}p^{i}\left\vert i\right\rangle \left\langle di\right\vert
\text{,} \label{edro}%
\end{equation}
respectively. By substituting the expression for $d\rho$ in Eq. (\ref{edro})
into Eq. (\ref{dbures}), $ds_{\text{Bures}}^{2}$ becomes%
\begin{equation}
ds_{\text{Bures}}^{2}=\frac{1}{4}\left[  \sum_{i=1}^{N}\frac{\left(
dp^{i}\right)  ^{2}}{p^{i}}+2\sum_{i\neq j}\frac{\left(  p^{i}-p^{j}\right)
^{2}}{p^{i}+p^{j}}\left\vert \left\langle i|dj\right\rangle \right\vert
^{2}\right]  \text{.} \label{dburess}%
\end{equation}
Observe that, modulo an irrelevant constant factor, $ds_{\text{Bures}}^{2}$ in
Eq. (\ref{dburess}) and $ds_{\text{DO}}^{2}$ in Eq. (\ref{qle}) are identical.
Similarly, note that when $\left[  \rho\text{, }d\rho\right]  \overset
{\text{def}}{=}\rho d\rho-d\rho\rho=0$, the Bures metric essentially becomes
the Fisher-Rao information metric. Furthermore, for pure states, $\rho
\overset{\text{def}}{=}\left\vert \psi\right\rangle \left\langle
\psi\right\vert $ with $d\rho\overset{\text{def}}{=}\left\vert d\psi
\right\rangle \left\langle \psi\right\vert +\left\vert \psi\right\rangle
\left\langle d\psi\right\vert $, the Bures metric becomes the Fubini-Study
metric,%
\begin{equation}
ds_{\text{Bures}}^{2}=\sum_{i\in\ker\left(  \rho\right)  }\left\vert
\left\langle d\psi|i\right\rangle \right\vert ^{2}=\left\langle d\psi|\left(
1-\left\vert \psi\right\rangle \left\langle \psi\right\vert \right)
|d\psi\right\rangle =\left\langle d\psi|d\psi\right\rangle -\left\langle
d\psi|\psi\right\rangle \left\langle \psi|d\psi\right\rangle =ds_{\text{FS}%
}^{2}\text{.}%
\end{equation}
For a more detailed presentation of this material, we refer to Refs.
\cite{sam, fuchs}. As a final remark, we point out that the distinguishability
of mixed density operators can be quantified in terms of several metrics
within the information geometric framework. For further details on this
specific issue, we refer to Ref. \cite{amari,karol,petz}.

\section{Geodesic paths in the projective space}

In this section, after pointing out our working assumptions, we use methods of
variational calculus to extremize the action functional expressed in terms of
the infinitesimal Fubini-Study line element. The extremization procedure leads
to determination of the geodesic paths followed by the quantum-mechanical
probability amplitudes of pure quantum states.

\subsection{Variance of the phase changes}

Recall that given two normalized pure states $\left\vert \psi\right\rangle $
and $\left\vert \psi^{\prime}\right\rangle $ as defined in Eq. (\ref{explicit}%
), the Fubini-Study metric becomes,%
\begin{equation}
ds_{\text{FS}}^{2}=\frac{1}{4}\left\{  \sum_{m=1}^{N}\frac{\dot{p}_{m}^{2}%
}{p_{m}}+4\left[  \sum_{m=1}^{N}p_{m}\dot{\phi}_{m}^{2}-\left(  \sum_{m=1}%
^{N}p_{m}\dot{\phi}_{m}\right)  ^{2}\right]  \right\}  d\theta^{2}\text{.}
\label{dfs}%
\end{equation}
In terms of the variance of phase changes $\sigma_{\dot{\phi}}^{2}$,%
\begin{equation}
\sigma_{\dot{\phi}}^{2}\overset{\text{def}}{=}\sum_{m=1}^{N}p_{m}\dot{\phi
}_{m}^{2}-\left(  \sum_{m=1}^{N}p_{m}\dot{\phi}_{m}\right)  ^{2}\text{,}
\label{variance1}%
\end{equation}
with $0\leq\sigma_{\dot{\phi}}^{2}$, the metric in\ Eq. (\ref{dfs}) becomes%
\begin{equation}
ds_{\text{FS}}^{2}=\frac{1}{4}\left[  \sum_{m=1}^{N}\frac{\dot{p}_{m}^{2}%
}{p_{m}}+4\sigma_{\dot{\phi}}^{2}\right]  d\theta^{2}\text{.}%
\end{equation}
Following a remark made in Ref. \cite{brau94}, we point out that a suitable
choice of an orthonormal basis $\left\{  \left\vert k\right\rangle \right\}  $
makes $\sigma_{\dot{\phi}}^{2}$ equal to zero. Specifically, the condition
that must be satisfied by $\left\{  \left\vert k\right\rangle \right\}  $ is
that for any $1\leq k\leq N$,%
\begin{equation}
p_{k}\left(  d\phi_{k}-\sum_{j=1}^{N}p_{j}d\phi_{j}\right)  =0\text{.}
\label{condition}%
\end{equation}
Indeed, after some simple algebraic manipulations,\ Eq. (\ref{condition}) can
be written as,%
\begin{equation}
\sum_{k=1}^{N}p_{k}\left(  d\phi_{k}\right)  ^{2}-\left(  \sum_{k=1}^{N}%
p_{k}d\phi_{k}\right)  ^{2}=0\text{,}%
\end{equation}
that is, using Eq. (\ref{variance1}), $\sigma_{\dot{\phi}}^{2}=0$. In
particular, any basis $\left\{  \left\vert k\right\rangle \right\}  $ that
satisfies Eq. (\ref{condition}) is such that its basis vectors also satisfy
the relation,%
\begin{equation}
\operatorname{Im}\left(  \left\langle k|\psi\right\rangle \left\langle
k|d\psi_{\bot}\right\rangle \right)  =0\text{,} \label{vrel}%
\end{equation}
with $\left\vert \psi\right\rangle $ and $\left\vert d\psi_{\bot}\right\rangle
$ given in Eqs. (\ref{explicit}) and (\ref{defi}), respectively. To verify the
relation in Eq. (\ref{vrel}), recall that $\left\vert d\psi\right\rangle
\overset{\text{def}}{=}\left\vert \psi^{\prime}\right\rangle -\left\vert
\psi\right\rangle $, $\left\vert d\psi_{\bot}\right\rangle \overset
{\text{def}}{=}\left\vert d\psi\right\rangle -\left\vert \psi\right\rangle
\left\langle \psi|d\psi\right\rangle $, where $\left\vert \psi\right\rangle $
and $\left\vert \psi^{\prime}\right\rangle $ can be written as%
\begin{equation}
\left\vert \psi\right\rangle \overset{\text{def}}{=}\sum_{j=1}^{N}\sqrt{p_{j}%
}e^{-i\phi_{j}}\left\vert j\right\rangle \text{, }%
\end{equation}
and,%
\begin{equation}
\left\vert \psi^{\prime}\right\rangle \overset{\text{def}}{=}\sum_{j=1}%
^{N}\sqrt{p_{j}+dp_{j}}e^{-i\left(  \phi_{j}+d\phi_{j}\right)  }\left\vert
j\right\rangle \text{, \ }%
\end{equation}
respectively. Using these previous four relations, $\left\langle
\psi|k\right\rangle \left\langle k|d\psi_{\bot}\right\rangle $ becomes%
\begin{equation}
\left\langle \psi|k\right\rangle \left\langle k|d\psi_{\bot}\right\rangle
=\sqrt{p_{k}}e^{-i\phi_{k}}\left[  \left\langle k|\psi^{\prime}\right\rangle
-\left\langle k|\psi\right\rangle \left\langle \psi|\psi^{\prime}\right\rangle
\right]  =\sqrt{p_{k}}e^{-i\phi_{k}}\left\langle k|\psi^{\prime}\right\rangle
-\sqrt{p_{k}}e^{-i\phi_{k}}\left\langle k|\psi\right\rangle \left\langle
\psi|\psi^{\prime}\right\rangle \text{.} \label{you1}%
\end{equation}
Observe that $\left\langle \psi|k\right\rangle =\sqrt{p_{k}}e^{-i\phi_{k}}$,
$\left\langle k|\psi\right\rangle =\left\langle \psi|k\right\rangle ^{\ast
}=\sqrt{p_{k}}e^{i\phi_{k}}$ and, limiting our attention to the second order
expansion of $\left\vert \psi^{\prime}\right\rangle $ with respect to
$d\phi_{k}$ and $dp_{k}$, the expressions for $\left\langle \psi|\psi^{\prime
}\right\rangle $ and $\left\langle k|\psi^{\prime}\right\rangle $ in Eq.
(\ref{you1}) become%
\begin{equation}
\left\langle \psi|\psi^{\prime}\right\rangle =1-\frac{1}{8}\sum_{j=1}^{N}%
\frac{\left(  dp_{j}\right)  ^{2}}{p_{j}}+i\sum_{j=1}^{N}p_{j}d\phi_{j}%
+\frac{i}{2}\sum_{j=1}^{N}dp_{j}d\phi_{j}-\frac{1}{2}\sum_{j=1}^{N}%
p_{j}\left(  d\phi_{j}\right)  ^{2}\text{,} \label{you22}%
\end{equation}
and,%
\begin{equation}
\left\langle k|\psi^{\prime}\right\rangle =\sqrt{p_{k}}e^{i\phi_{k}}\left[
1+i\left(  d\phi_{k}\right)  -\frac{1}{2}\left(  d\phi_{k}\right)  ^{2}%
+\frac{1}{2}\frac{dp_{k}}{p_{k}}+\frac{i}{2}\frac{dp_{k}d\phi_{k}}{p_{k}%
}-\frac{1}{8}\frac{\left(  dp_{k}\right)  ^{2}}{p_{k}^{2}}\right]  \text{,}
\label{you23}%
\end{equation}
respectively. Substituting Eqs. (\ref{you22}) and (\ref{you23}) into Eq.
(\ref{you1}), after some algebra, we obtain Eq. (\ref{vrel}). In conclusion,
it is always possible to assume $\sigma_{\dot{\phi}}^{2}=0$ given an
appropriate choice of basis $\left\{  \left\vert k\right\rangle \right\}  $.
In what follows, we assume to be working\textbf{ }under such a condition.

\subsection{Extremizing the action functional}

For the sake of convenience, recall that the relation between the
Wigner-Yanase and the Fubini-Study infinitesimal line elements is
$ds_{\text{WY}}^{2}=4ds_{\text{FS}}^{2}$, where $ds_{\text{FS}}^{2}$ is given
by%
\begin{equation}
ds_{\text{FS}}^{2}\overset{\text{def}}{=}\frac{1}{4}\left\{  \sum_{k=1}%
^{N}\frac{\dot{p}_{k}^{2}}{p_{k}}+4\left[  \sum_{k=1}^{N}\dot{\phi}_{k}%
^{2}p_{k}-\left(  \sum_{k=1}^{N}\dot{\phi}_{k}p_{k}\right)  ^{2}\right]
\right\}  d\theta^{2}\text{,} \label{fubini}%
\end{equation}
with $\dot{p}_{k}$ and $\dot{\phi}_{k}$ defined in Eq. (\ref{okoggi}). Observe
that,%
\begin{equation}
\sum_{k=1}^{N}\dot{\phi}_{k}^{2}p_{k}-\left(  \sum_{k=1}^{N}\dot{\phi}%
_{k}p_{k}\right)  ^{2}=\left\langle \dot{\phi}^{2}\right\rangle -\left\langle
\dot{\phi}\right\rangle ^{2}=\sigma_{\dot{\phi}}^{2}\text{,} \label{variance}%
\end{equation}
where $\left\langle \cdot\right\rangle $ denotes the averaging operation and
$\phi\overset{\text{def}}{=}\left(  \phi_{1}\text{,..., }\phi_{N}\right)  $.
Therefore, using Eq. (\ref{variance}), Eq. (\ref{fubini}) can be rewritten as%
\begin{equation}
ds_{\text{FS}}^{2}=\frac{1}{4}\left\{  \mathcal{F}\left(  \theta\right)
+4\sigma_{\dot{\phi}}^{2}\right\}  d\theta^{2}\text{,}%
\end{equation}
where $\mathcal{F}\left(  \theta\right)  $ denotes the Fisher information
function defined as,%
\begin{equation}
\mathcal{F}\left(  \theta\right)  \overset{\text{def}}{=}\sum_{k=1}^{N}%
\frac{\dot{p}_{k}^{2}}{p_{k}}\text{.} \label{frf}%
\end{equation}
As pointed out earlier, we assume $\sigma_{\dot{\phi}}^{2}=0$. Then, the
action functional to consider is given by%
\begin{equation}
\mathcal{S}\overset{\text{def}}{=}\int ds_{\text{FS}}=\int\sqrt{ds_{\text{FS}%
}^{2}}=\frac{1}{2}\int\mathcal{F}^{\frac{1}{2}}\left(  \theta\right)
d\theta\text{,}%
\end{equation}
that is, more formally,%
\begin{equation}
\mathcal{S}\left[  p\right]  \overset{\text{def}}{=}\int\mathcal{L}\left(
\dot{p}\text{, }p\text{, }\theta\right)  d\theta\text{,} \label{action}%
\end{equation}
where $p\overset{\text{def}}{=}\left(  p_{1}\text{,..., }p_{N}\right)  $ and
$\mathcal{L}\left(  \dot{p}\text{, }p\text{, }\theta\right)  $ denotes the
Lagrangian-like function defined as,%
\begin{equation}
\mathcal{L}\left(  \dot{p}\text{, }p\text{, }\theta\right)  \overset
{\text{def}}{=}\frac{1}{2}\mathcal{F}^{\frac{1}{2}}\left(  \theta\right)
\text{,} \label{lagrangian}%
\end{equation}
with $\mathcal{F}\left(  \theta\right)  $ defined in Eq. (\ref{frf}). We wish
to determine the probability paths $p\overset{\text{def}}{=}\left(
p_{1}\text{,..., }p_{N}\right)  $ with $p_{k}=p_{k}\left(  \theta\right)  $
for any $1\leq k\leq N$ that make the action functional $\mathcal{S}\left[
p\right]  $ in Eq. (\ref{action}) stationary subject to the conservation of
the probability condition,%
\begin{equation}
\sum_{k=1}^{N}p_{k}=1\text{.} \label{nc}%
\end{equation}
Generally speaking, an action functional $\mathcal{S}\left[  p\right]
\overset{\text{def}}{=}\int\mathcal{L}\left(  \dot{p}\text{, }p\text{, }%
\theta\right)  d\theta$ has a stationary value if $\delta\mathcal{S}=0$. It
happens that for any $\bar{k}$ with $1\leq\bar{k}\leq N$, we have%
\begin{align}
\delta\mathcal{S}  &  =\int\left(  \frac{\partial\mathcal{L}}{\partial\dot
{p}_{\bar{k}}}\delta\dot{p}_{\bar{k}}+\frac{\partial\mathcal{L}}{\partial
p_{\bar{k}}}\delta p_{\bar{k}}\right)  d\theta\nonumber\\
& \nonumber\\
&  =\int\left(  \frac{\partial\mathcal{L}}{\partial\dot{p}_{\bar{k}}}%
\frac{d\left(  \delta p_{\bar{k}}\right)  }{d\theta}+\frac{\partial
\mathcal{L}}{\partial p_{\bar{k}}}\delta p_{\bar{k}}\right)  d\theta
\nonumber\\
& \nonumber\\
&  =\int\frac{\partial\mathcal{L}}{\partial\dot{p}_{\bar{k}}}\frac{d\left(
\delta p_{\bar{k}}\right)  }{d\theta}d\theta+\int\frac{\partial\mathcal{L}%
}{\partial p_{\bar{k}}}\delta p_{\bar{k}}d\theta\text{.} \label{variation}%
\end{align}
Integrating by parts the first term in the last line of Eq. (\ref{variation}),
we find%
\begin{equation}
\delta\mathcal{S}=\frac{\partial\mathcal{L}}{\partial\dot{p}_{\bar{k}}}\delta
p_{\bar{k}}-\int\frac{d}{d\theta}\left(  \frac{\partial\mathcal{L}}%
{\partial\dot{p}_{\bar{k}}}\right)  \delta p_{\bar{k}}d\theta+\int
\frac{\partial\mathcal{L}}{\partial p_{\bar{k}}}\delta p_{\bar{k}}%
d\theta\text{.}%
\end{equation}
We point out that, in the variational calculus scheme being considered here,
only the probability paths $p_{\bar{k}}\left(  \theta\right)  $ are being
varied while the endpoints are being kept fix, that is, $\delta p_{\bar{k}%
}\left(  \theta_{\text{i}}\right)  =\delta p_{\bar{k}}\left(  \theta
_{\text{f}}\right)  =0$. Therefore, the condition $\delta\mathcal{S}=0$
becomes%
\begin{equation}
\int\left[  \frac{d}{d\theta}\left(  \frac{\partial\mathcal{L}}{\partial
\dot{p}_{\bar{k}}}\right)  -\frac{\partial\mathcal{L}}{\partial p_{\bar{k}}%
}\right]  \delta p_{\bar{k}}d\theta=0\text{.} \label{almost}%
\end{equation}
Since Eq. (\ref{almost}) must be satisfied for any small change $\delta
p_{\bar{k}}$, the condition $\delta\mathcal{S}=0$ leads to the so-called
Euler-Lagrange differential equations:%
\begin{equation}
\frac{d}{d\theta}\left(  \frac{\partial\mathcal{L}}{\partial\dot{p}_{\bar{k}}%
}\right)  -\frac{\partial\mathcal{L}}{\partial p_{\bar{k}}}=0\text{.}%
\end{equation}
Returning to our specific problem, we wish to find the stationary value of the
action functional,%
\begin{equation}
\mathcal{S}\left[  p\right]  =\int\left[  \mathcal{L}\left(  \dot{p}\text{,
}p\text{, }\theta\right)  -\lambda\left(  \sum_{k=1}^{N}p_{k}-1\right)
\right]  d\theta\text{,} \label{go}%
\end{equation}
where $\lambda$ in Eq. (\ref{go}) is the Lagrange multiplier coefficient and
$\mathcal{L}\left(  \dot{p}\text{, }p\text{, }\theta\right)  $ is the
Lagrangian-like function given in\ Eq. (\ref{lagrangian}). Consider the
following change of variables \cite{wootters81},%
\begin{equation}
p_{k}\left(  \theta\right)  \rightarrow q_{k}\left(  \theta\right)  \text{,
with }p_{k}\left(  \theta\right)  \overset{\text{def}}{=}q_{k}^{2}\left(
\theta\right)  \text{. }%
\end{equation}
In terms of the probability amplitude variables $q_{k}\left(  \theta\right)
$, Eqs. (\ref{frf}) and (\ref{nc}) become%
\begin{equation}
\mathcal{F}\left(  \theta\right)  =4\sum_{k=1}^{N}\dot{q}_{k}^{2}\text{,}
\label{frf2}%
\end{equation}
and,%
\begin{equation}
\sum_{k=1}^{N}q_{k}^{2}=1\text{,} \label{nc2}%
\end{equation}
respectively. Using Eqs. (\ref{frf2}) and (\ref{nc2}), the action functional
in Eq. (\ref{go}) becomes%
\begin{equation}
\mathcal{S}_{\text{new}}\left[  q\right]  =\int\mathcal{L}_{\text{new}}\left(
\dot{q}\text{, }q\text{, }\theta\right)  d\theta\text{,}%
\end{equation}
where $q\overset{\text{def}}{=}\left(  q_{1}\text{,..., }q_{N}\right)  $, and%
\begin{equation}
\mathcal{L}_{\text{new}}\left(  \dot{q}\text{, }q\text{, }\theta\right)
\overset{\text{def}}{=}\left(  \sum_{k=1}^{N}\dot{q}_{k}^{2}\right)
^{\frac{1}{2}}-\lambda\left(  \sum_{k=1}^{N}q_{k}^{2}-1\right)  \text{.}%
\end{equation}
Following the line of reasoning outlined before, we find that%
\begin{equation}
\delta\mathcal{S}_{\text{new}}=\frac{\delta\mathcal{S}_{\text{new}}}{\delta
q_{\bar{k}}}\delta q_{\bar{k}}=0\text{, }\forall\text{ }1\leq\bar{k}\leq
N\text{, }%
\end{equation}
leads to the following Euler-Lagrange differential equations,%
\begin{equation}
\frac{d}{d\theta}\left(  \frac{\partial\mathcal{L}_{\text{new}}}{\partial
\dot{q}_{\bar{k}}}\right)  -\frac{\partial\mathcal{L}_{\text{new}}}{\partial
q_{\bar{k}}}=0\text{.} \label{eln}%
\end{equation}
A straightforward computation yields the following three equalities,%
\begin{align}
\frac{\partial\mathcal{L}_{\text{new}}}{\partial\dot{q}_{\bar{k}}}  &
=\frac{\dot{q}_{\bar{k}}}{\left(  \sum_{k=1}^{N}\dot{q}_{k}^{2}\right)
^{\frac{1}{2}}}\text{, }\nonumber\\
& \nonumber\\
\text{ }\frac{d}{d\theta}\left(  \frac{\partial\mathcal{L}_{\text{new}}%
}{\partial\dot{q}_{\bar{k}}}\right)   &  =\frac{\ddot{q}_{\bar{k}}}{\left(
\sum_{k=1}^{N}\dot{q}_{k}^{2}\right)  ^{\frac{1}{2}}}\text{ }-\frac{\dot
{q}_{\bar{k}}^{2}\ddot{q}_{\bar{k}}}{\left(  \sum_{k=1}^{N}\dot{q}_{k}%
^{2}\right)  ^{\frac{3}{2}}}\text{,}\nonumber\\
& \nonumber\\
\frac{\partial\mathcal{L}_{\text{new}}}{\partial q_{\bar{k}}}  &  =-2\lambda
q_{\bar{k}}\text{.} \label{computation}%
\end{align}
Employing the three relations in Eq. (\ref{computation}), the Euler-Lagrange
equations in\ Eq. (\ref{eln}) become%
\begin{equation}
\frac{\ddot{q}_{\bar{k}}}{\left(  \sum_{k=1}^{N}\dot{q}_{k}^{2}\right)
^{\frac{1}{2}}}\text{ }-\frac{\dot{q}_{\bar{k}}^{2}\ddot{q}_{\bar{k}}}{\left(
\sum_{k=1}^{N}\dot{q}_{k}^{2}\right)  ^{\frac{3}{2}}}+2\lambda q_{\bar{k}%
}=0\text{,}%
\end{equation}
that is,%
\begin{equation}
\ddot{q}_{\bar{k}}\text{ }-\frac{\dot{q}_{\bar{k}}^{2}\ddot{q}_{\bar{k}}}%
{\sum_{k=1}^{N}\dot{q}_{k}^{2}}+2\lambda\left(  \sum_{k=1}^{N}\dot{q}_{k}%
^{2}\right)  ^{\frac{1}{2}}q_{\bar{k}}=0\text{.} \label{el3}%
\end{equation}
Observe that in terms of the Lagrangian-like function $\mathcal{L}$ defined in
Eq. (\ref{lagrangian}) expressed in terms of the probability amplitudes
$q_{k}$, we find that $\mathcal{L}\left(  \theta\right)  $ and $\mathcal{\dot
{L}}\left(  \theta\right)  /\mathcal{L}\left(  \theta\right)  $ are given by,
\begin{equation}
\mathcal{L}\left(  \theta\right)  =\frac{1}{2}\mathcal{F}^{\frac{1}{2}}\left(
\theta\right)  =\left(  \sum_{k=1}^{N}\dot{q}_{k}^{2}\right)  ^{\frac{1}{2}%
}\text{, and }\frac{\mathcal{\dot{L}}\left(  \theta\right)  }{\mathcal{L}%
\left(  \theta\right)  }=\frac{\dot{q}_{\bar{k}}\ddot{q}_{\bar{k}}}{\sum
_{k=1}^{N}\dot{q}_{k}^{2}}\text{,} \label{buono}%
\end{equation}
respectively. Then, using the equalities in\ Eq. (\ref{buono}), the
Euler-Lagrange equations expressed in Eq. (\ref{el3}) become%
\begin{equation}
\ddot{q}_{\bar{k}}\text{ }-\frac{\mathcal{\dot{L}}\left(  \theta\right)
}{\mathcal{L}\left(  \theta\right)  }\dot{q}_{\bar{k}}+2\lambda_{\text{FS}%
}\mathcal{L}\left(  \theta\right)  q_{\bar{k}}=0\text{,} \label{finalmente}%
\end{equation}
where $\lambda_{\text{FS}}$ is the Lagrange multiplier coefficient obtained
within the framework of the Fubini-Study metric. For the sake of completeness,
we remark that if we had used the Wigner-Yanase metric, Eq. (\ref{finalmente})
would have been written as%
\begin{equation}
\ddot{q}_{\bar{k}}\text{ }-\frac{\mathcal{\dot{L}}\left(  \theta\right)
}{\mathcal{L}\left(  \theta\right)  }\dot{q}_{\bar{k}}+\frac{\lambda
_{\text{WY}}}{2}\mathcal{L}\left(  \theta\right)  q_{\bar{k}}=0\text{.}
\label{finalmente2}%
\end{equation}
The rescaling of the Lagrange multiplier coefficient in transitioning from the
Fubini-Study to the Wigner-Yanase cases occurs in order to satisfy the
conservation of probability condition in both scenarios. Finally, recalling
that
\begin{equation}
\mathcal{L}_{\text{FS}}=\frac{1}{2}\mathcal{F}^{\frac{1}{2}}\text{, and
}\mathcal{L}_{\text{WY}}=\mathcal{F}^{\frac{1}{2}}\text{,}%
\end{equation}
that is, $\mathcal{F}=4\mathcal{L}_{\text{FS}}^{2}=\mathcal{L}_{\text{WY}}%
^{2}$, in terms of the Fisher information function, Eqs. (\ref{finalmente})
and (\ref{finalmente2}) become%
\begin{equation}
\ddot{q}_{\bar{k}}\text{ }-\frac{1}{2}\frac{\mathcal{\dot{F}}\left(
\theta\right)  }{\mathcal{F}\left(  \theta\right)  }\dot{q}_{\bar{k}}%
+\lambda_{\text{FS}}\mathcal{F}^{\frac{1}{2}}\left(  \theta\right)  q_{\bar
{k}}=0\text{,} \label{fsode}%
\end{equation}
and,%
\begin{equation}
\ddot{q}_{\bar{k}}\text{ }-\frac{1}{2}\frac{\mathcal{\dot{F}}\left(
\theta\right)  }{\mathcal{F}\left(  \theta\right)  }\dot{q}_{\bar{k}}%
+\frac{\lambda_{\text{WY}}}{2}\mathcal{F}^{\frac{1}{2}}\left(  \theta\right)
q_{\bar{k}}=0\text{,} \label{wyode}%
\end{equation}
respectively. In general, for each $\bar{k}$ with $1\leq\bar{k}\leq N$, the
integration of the previous second-order $N$ ordinary differential equations
(ODEs) leads to a formal expression of $q_{\bar{k}}\left(  \theta\right)  $.
Specifically, each $q_{\bar{k}}\left(  \theta\right)  $ is the superposition
of two linearly independent solutions of the ODEs expressed in terms of two
real constants of integration $c_{\bar{k}}^{\left(  1\right)  }$ and
$c_{\bar{k}}^{\left(  2\right)  }$. In particular, the formal expressions of
such independent solutions appear in terms of the Lagrange multiplier
$\lambda_{\text{FS}}$ and depend on the particular functional dependence of
the Fisher information function on the parameter $\theta$. Therefore, in
principle, the exact expression of these independent solutions requires one to
express the Lagrange multiplier in terms of the characteristic parameters that
specify the Fisher information function by imposing that $\frac{1}{4}F\left(
\theta\right)  $ equals the sum of $\dot{q}_{\bar{k}}^{2}\left(
\theta\right)  $ with $1\leq\bar{k}\leq N$. Finally, since $q_{\bar{k}}%
^{2}\left(  \theta\right)  $ are probabilities, the $2N$ integration constants
$c_{\bar{k}}^{\left(  1\right)  }$ and $c_{\bar{k}}^{\left(  2\right)  }$ have
to be chosen in such a manner that $q_{\bar{k}}^{2}\left(  \theta\right)  $
add up to unity and $0\leq q_{\bar{k}}^{2}\left(  \theta\right)  \leq1$ for
any $1\leq\bar{k}\leq N$. In what follows, we shall take into consideration
the integration of Eq. (\ref{fsode}) for various functional forms of the
Fisher information function.

\section{Illustrative Examples}

In this section, we determine the geodesic paths followed by the
quantum-mechanical probability amplitudes of pure quantum states for three
distinct functional forms of the Fisher information function: constant Fisher
information, exponential decay, and power-law decay.

\subsection{Example one: Constant Fisher information}

In the working assumption that $\mathcal{F}\left(  \theta\right)  $ takes a
constant value $\mathcal{F}_{0}$, Eq. (\ref{fsode}) describes a simple
harmonic oscillator,%
\begin{equation}
\ddot{q}_{\bar{k}}+\lambda_{\text{FS}}\mathcal{F}_{0}^{\frac{1}{2}}q_{\bar{k}%
}=0\text{.} \label{fsode1}%
\end{equation}
Integration of Eq. (\ref{fsode1}) leads to the following general solution for
the geodesic path of quantum-mechanical probability amplitudes $q_{\bar{k}%
}\left(  \theta\right)  $,%
\begin{equation}
q_{\bar{k}}\left(  \theta\right)  =c_{\bar{k}}^{\left(  1\right)  }\cos\left(
\mathcal{F}_{0}^{\frac{1}{4}}\sqrt{\lambda_{\text{FS}}}\theta\right)
+c_{\bar{k}}^{\left(  2\right)  }\sin\left(  \mathcal{F}_{0}^{\frac{1}{4}%
}\sqrt{\lambda_{\text{FS}}}\theta\right)  \text{,}%
\end{equation}
where $c_{\bar{k}}^{\left(  1\right)  }$ and $c_{\bar{k}}^{\left(  2\right)
}$ are two integration constants. Therefore, assuming for the sake of
clarity\textbf{ }that $\bar{k}=1$, $2$, probabilities $p_{1}\left(
\theta\right)  $ and $p_{2}\left(  \theta\right)  \overset{\text{def}}%
{=}1-p_{1}\left(  \theta\right)  $ can be written as%
\begin{equation}
p_{1}\left(  \theta\right)  =\cos^{2}\left(  \mathcal{F}_{0}^{\frac{1}{4}%
}\sqrt{\lambda_{\text{FS}}}\theta\right)  \text{, and }p_{2}\left(
\theta\right)  =\sin^{2}\left(  \mathcal{F}_{0}^{\frac{1}{4}}\sqrt
{\lambda_{\text{FS}}}\theta\right)  \text{,} \label{pexplicit}%
\end{equation}
respectively. The value of the Lagrange multiplier coefficient $\lambda
_{\text{FS}}$ in Eq. (\ref{pexplicit}) can be explicitly obtained by requiring
that,%
\begin{equation}
\frac{\dot{p}_{1}^{2}\left(  \theta\right)  }{p_{1}\left(  \theta\right)
}+\frac{\dot{p}_{2}^{2}\left(  \theta\right)  }{p_{2}\left(  \theta\right)
}=\mathcal{F}_{0}\text{.} \label{conditionf}%
\end{equation}
By substituting Eq. (\ref{pexplicit}) into Eq. (\ref{conditionf}), we obtain%
\begin{equation}
\lambda_{\text{FS}}=\frac{1}{4}\mathcal{F}_{0}^{\frac{1}{2}}\text{.}%
\end{equation}
\begin{figure}[ptb]
\centering
\includegraphics[width=0.35\textwidth] {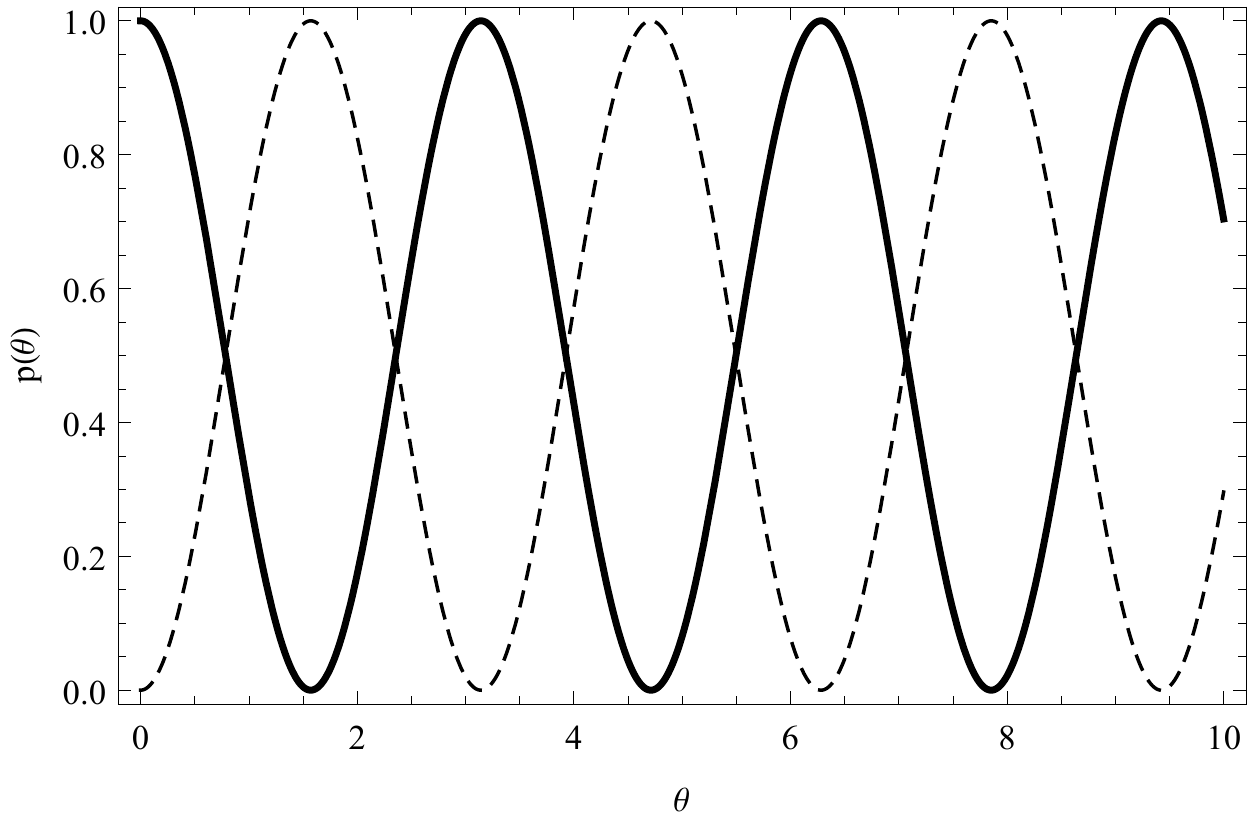}\caption{Oscillatory behavior of
the success (dotted) and failure (solid) probabilities in the case of constant
Fisher information.}%
\label{fig1}%
\end{figure}Observe that in the case of the analog counterpart of Grover's
quantum search algorithm $\mathcal{F}_{0}=4$ and, thus, $\lambda_{\text{FS}%
}=1/2$ and $\lambda_{\text{WY}}=1$. Therefore, the failure and success
probabilities are given by $p_{1}\left(  \theta\right)  =\cos^{2}\left(
\theta\right)  $ and $p_{2}\left(  \theta\right)  =\sin^{2}\left(
\theta\right)  $, respectively. In Fig. $1$, we observe an oscillatory
behavior of the success (dotted) and failure (solid) probabilities in the case
of constant Fisher information.

\subsection{Example two: Exponential decay}

In this subsection, we assume that the Fisher information function is a
monotonically decreasing function that exhibits exponentially decaying
behavior,%
\begin{equation}
\mathcal{F}\left(  \theta\right)  \overset{\text{def}}{=}\mathcal{F}%
_{0}e^{-\xi\theta}\text{,}%
\end{equation}
with $\mathcal{F}_{0}$ and $\xi$ being\textbf{ }positive real constant
coefficients. In this working assumption, Eq. (\ref{fsode}) describes the
equation of an aging spring in the presence of damping,%
\begin{equation}
\ddot{q}_{\bar{k}}+\frac{\xi}{2}\dot{q}_{\bar{k}}+\lambda_{\text{FS}%
}\mathcal{F}_{0}^{\frac{1}{2}}e^{-\frac{\xi}{2}\theta}q_{\bar{k}}=0\text{,}
\label{fsode2}%
\end{equation}
where $\dot{q}_{\bar{k}}\overset{\text{def}}{=}dq_{\bar{k}}/d\theta$. Equation
(\ref{fsode2}) can be analytically integrated, and a closed form solution can
be found.

In what follows, we consider the classical second order ordinary differential
equation that describes an aging spring with damping,%
\begin{equation}
m\ddot{x}+b\dot{x}+ke^{-\eta t}x=0\text{,} \label{derv1}%
\end{equation}
that is,%
\begin{equation}
\ddot{x}+\frac{b}{m}\dot{x}+\frac{k}{m}e^{-\eta t}x=0\text{.} \label{derv2}%
\end{equation}
In Eq. (\ref{derv2}), $m>0$ is the mass, $k>0$ is the value of the spring
constant at $t=0$, $b>0$ is the constant damping coefficient, $\eta
\overset{\text{def}}{=}-\frac{1}{k\left(  t\right)  }\frac{d\left[  k\left(
t\right)  \right]  }{dt}\in%
\mathbb{R}
_{+}\backslash\left\{  0\right\}  $, and $\dot{x}\overset{\text{def}}{=}%
dx/dt$. Equations (\ref{fsode2}) and (\ref{derv2}) are essentially identical
once we impose that,%
\begin{equation}
\theta=t\text{, }\xi=\frac{2b}{m}=2\eta\text{, and }\lambda_{\text{FS}%
}\mathcal{F}_{0}^{\frac{1}{2}}=\frac{k}{m}\text{.} \label{corri}%
\end{equation}
To integrate Eq. (\ref{derv1}), we employ two convenient mathematical tricks.
First, we make a change of variables,%
\begin{equation}
x\left(  t\right)  \rightarrow y\left(  t\right)  :x\left(  t\right)
\overset{\text{def}}{=}y\left(  t\right)  e^{-\frac{b}{2m}t}\text{.}
\label{cov2}%
\end{equation}
From Eq. (\ref{cov2}), we get%
\begin{equation}
\dot{x}=\dot{y}e^{-\frac{b}{2m}t}-\frac{b}{2m}ye^{-\frac{b}{2m}t}\text{, and
}\ddot{x}=\ddot{y}e^{-\frac{b}{2m}t}-\frac{b}{m}\dot{y}e^{-\frac{b}{2m}%
t}+\frac{b^{2}}{4m^{2}}ye^{-\frac{b}{2m}t}\text{.} \label{derv}%
\end{equation}
Using the two relations in Eqs. (\ref{derv}), Eq. (\ref{derv1}) becomes%
\begin{equation}
m\left(  \ddot{y}-\frac{b}{m}\dot{y}+\frac{b^{2}}{4m^{2}}y\right)  +b\left(
\dot{y}-\frac{b}{2m}y\right)  +ke^{-\eta t}y=0\text{,}%
\end{equation}
that is,%
\begin{equation}
m\ddot{y}+\left(  ke^{-\eta t}-\frac{b^{2}}{4m}\right)  y=0\text{.}
\label{eq2}%
\end{equation}
At this point, let us consider the change of the independent variable%
\begin{equation}
t\rightarrow s\left(  t\right)  :s\left(  t\right)  \overset{\text{def}}%
{=}\alpha e^{\beta t}\text{,} \label{cov}%
\end{equation}
that is,%
\begin{equation}
t=\frac{1}{\beta}\log\left(  \frac{s}{\alpha}\right)  \text{,}%
\end{equation}
with $\alpha$ and $\beta$ being\textbf{ }real coefficients. From Eq.
(\ref{cov}), we obtain after some algebra,%
\begin{equation}
\frac{d}{dt}\overset{\text{def}}{=}\beta s\frac{d}{ds}\text{, and }\frac
{d^{2}}{dt^{2}}\overset{\text{def}}{=}\beta^{2}s\left(  \frac{d}{ds}%
+s\frac{d^{2}}{ds^{2}}\right)  \text{.} \label{sub2}%
\end{equation}
Using the relations in Eq. (\ref{sub2}), Eq. (\ref{eq2}) becomes%
\begin{equation}
s^{2}y^{\prime\prime}+sy^{\prime}+\frac{1}{m\beta^{2}}\left[  k\left(
\frac{s}{\alpha}\right)  ^{-\frac{\eta}{\beta}}-\frac{b^{2}}{4m}\right]
y=0\text{,} \label{beta}%
\end{equation}
where $y^{\prime}\overset{\text{def}}{=}dy/ds$. At this point, we impose that%
\begin{equation}
-\frac{\eta}{\beta}\overset{\text{def}}{=}2\text{, and }\frac{k}{m\beta
^{2}\alpha^{-\frac{\eta}{\beta}}}\overset{\text{def}}{=}1\text{,}%
\end{equation}
that is,%
\begin{equation}
\alpha\overset{\text{def}}{=}\frac{2}{\eta}\sqrt{\frac{k}{m}}\text{, and
}\beta\overset{\text{def}}{=}-\frac{1}{2}\eta\text{,} \label{alpha}%
\end{equation}
and, consequently,%
\begin{equation}
s\left(  t\right)  =\frac{2}{\eta}\sqrt{\frac{k}{m}}e^{-\frac{1}{2}\eta
t}\text{.}%
\end{equation}
Using the relations in Eq. (\ref{alpha}), the linear second-order differential
equation in Eq. (\ref{beta}) becomes%
\begin{equation}
s^{2}y^{\prime\prime}+sy^{\prime}+\left[  1-\left(  \frac{b}{m\eta}\right)
^{2}\right]  y=0\text{.} \label{bessell}%
\end{equation}
Equation (\ref{bessell}) is Bessel's equation of order $\frac{b}{m\eta}\geq0$
and its integration leads to the following general solution \cite{king03},%
\begin{equation}
y\left(  s\right)  =c_{1}\mathcal{J}_{+\frac{b}{m\eta}}\left(  s\right)
+c_{2}\mathcal{J}_{-\frac{b}{m\eta}}\left(  s\right)  \text{,}%
\end{equation}
that is, using Eqs. (\ref{cov2}), (\ref{cov}) and (\ref{alpha}),%
\begin{equation}
x(t)=c_{1}e^{-\frac{b}{2m}t}\mathcal{J}_{+\frac{b}{m\eta}}\left(  \frac
{2}{\eta}\sqrt{\frac{k}{m}}e^{-\frac{\eta}{2}t}\right)  +c_{2}e^{-\frac{b}%
{2m}t}\mathcal{J}_{-\frac{b}{m\eta}}\left(  \frac{2}{\eta}\sqrt{\frac{k}{m}%
}e^{-\frac{\eta}{2}t}\right)  \text{,} \label{solution}%
\end{equation}
where $c_{1}$ and $c_{2}$ are two real integration constants, and
$\mathcal{J}_{\upsilon}\left(  x\right)  $ denotes the Bessel function of the
first kind of order $\nu\geq0$ \cite{king03}. Finally, using Eqs.
(\ref{corri}) and (\ref{solution}), the geodesic path of the
quantum-mechanical probability amplitudes $q_{\bar{k}}\left(  \theta\right)  $
becomes,%
\begin{equation}
q_{\bar{k}}\left(  \theta\right)  =c_{\bar{k}}^{\left(  1\right)  }%
e^{-\frac{\xi}{4}\theta}\mathcal{J}_{+1}\left(  \frac{4}{\xi}\sqrt
{\lambda_{\text{FS}}}\mathcal{F}_{0}^{\frac{1}{4}}e^{-\frac{\xi}{4}\theta
}\right)  +c_{\bar{k}}^{\left(  2\right)  }e^{-\frac{\xi}{4}\theta}%
\mathcal{J}_{-1}\left(  \frac{4}{\xi}\sqrt{\lambda_{\text{FS}}}\mathcal{F}%
_{0}^{\frac{1}{4}}e^{-\frac{\xi}{4}\theta}\right)  \text{,}%
\end{equation}
where $c_{\bar{k}}^{\left(  1\right)  }$ and $c_{\bar{k}}^{\left(  2\right)
}$ are two real integration constants. In Fig. $2$, setting $\mathcal{F}%
_{0}=1$, $\xi=2$ and preserving the normalization constraint, we observe a
monotonic behavior of the success (dotted) and failure (solid) probabilities
in the case of exponential decay of the Fisher information.

\begin{figure}[ptb]
\centering
\includegraphics[width=0.35\textwidth] {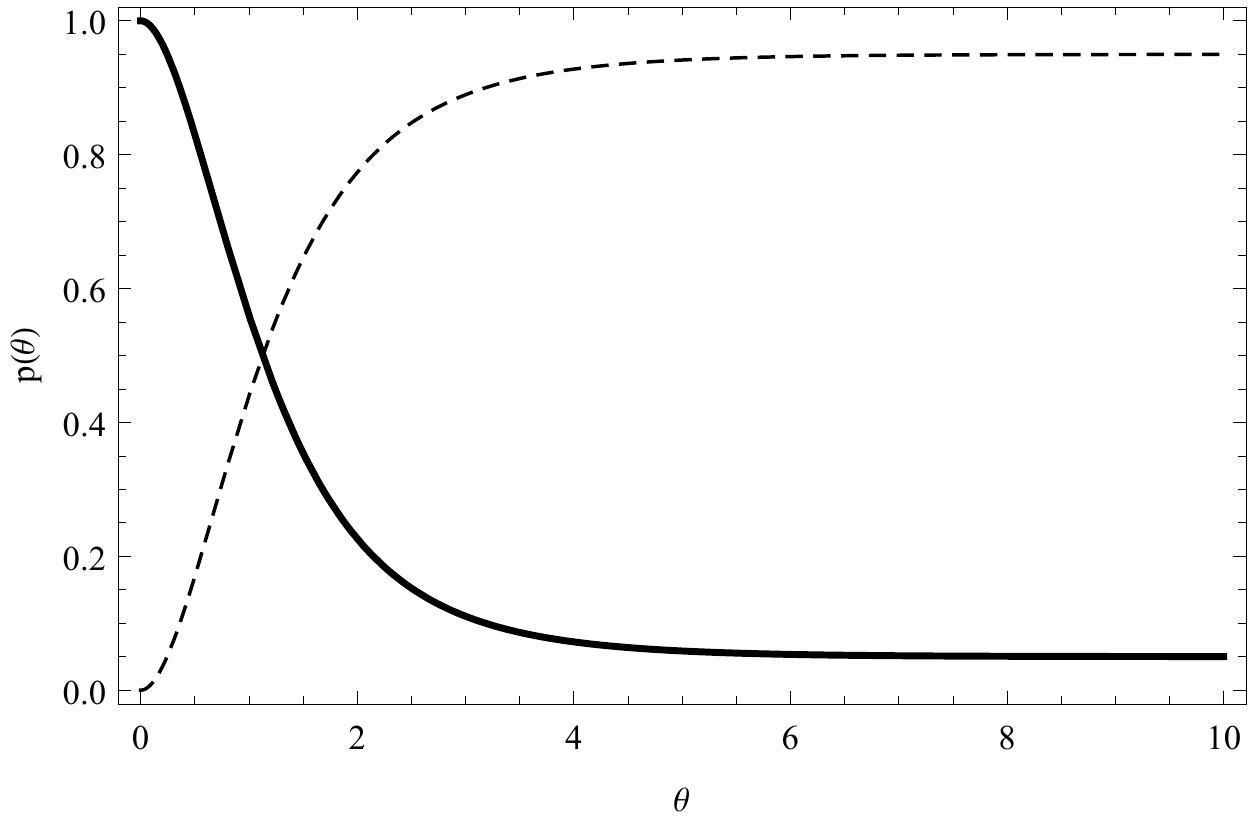}\caption{Monotonic behavior of
the success (dotted) and failure (solid) probabilities in the case of
exponential decay of the Fisher information.}%
\label{fig2}%
\end{figure}

\subsection{Example three: Power-law decay}

In this subsection, we assume that the Fisher information function is a
monotonically decreasing function that exhibits power-law decay behavior,%
\begin{equation}
\mathcal{F}\left(  \theta\right)  \overset{\text{def}}{=}\frac{\mathcal{F}%
_{0}}{\left(  1+\Omega\theta\right)  ^{n}}\text{,} \label{power}%
\end{equation}
where $\mathcal{F}_{0}$, $\Omega$, and $n\geq0$ are real constant
coefficients. Using Eq. (\ref{power}), Eq. (\ref{fsode}) becomes%
\begin{equation}
\ddot{q}_{\bar{k}}+\frac{n\Omega}{2}\frac{1}{1+\Omega\theta}\dot{q}_{\bar{k}%
}+\frac{\lambda_{\text{FS}}\mathcal{F}_{0}^{\frac{1}{2}}}{\left(
1+\Omega\theta\right)  ^{\frac{n}{2}}}q_{\bar{k}}=0\text{,} \label{powere}%
\end{equation}
where $\dot{q}_{\bar{k}}\overset{\text{def}}{=}dq_{\bar{k}}/d\theta$. Equation
(\ref{powere}) is a linear second-order ordinary differential equation with
varying coefficients. Its analytical integration is nontrivial for arbitrary
values of $n\geq0$. However, in what follows, we use a mathematical trick that
allows to deduce a closed form solution for Eq. (\ref{powere}) for a specific
choice of the constants $\Omega$ and $n$ in Eq. (\ref{power}). We proceed as follows.

Consider the second order linear differential equation with time-dependent
coefficients,%
\begin{equation}
\ddot{x}+p\left(  t\right)  \dot{x}+q\left(  t\right)  x=0\text{,}
\label{original}%
\end{equation}
where $\dot{x}\overset{\text{def}}{=}dx/dt$. Next, consider the following
change of independent variable,%
\begin{equation}
t\rightarrow s:s\overset{\text{def}}{=}f\left(  t\right)  \text{.}%
\end{equation}
After some algebra, we obtain%
\begin{equation}
\frac{dx}{dt}=\frac{ds}{dt}\frac{dx}{ds}\text{,} \label{firstd}%
\end{equation}
and,%
\begin{align}
\frac{d^{2}x}{dt^{2}}  &  =\frac{d}{dt}\left(  \frac{dx}{dt}\right)  =\frac
{d}{dt}\left(  \frac{ds}{dt}\frac{dx}{ds}\right)  =\frac{ds}{dt}\frac{d}%
{ds}\left(  \frac{ds}{dt}\frac{dx}{ds}\right) \nonumber\\
& \nonumber\\
&  =\frac{ds}{dt}\frac{d}{ds}\left(  \frac{ds}{dt}\right)  \frac{dx}{ds}%
+\frac{ds}{dt}\frac{ds}{dt}\frac{d}{ds}\left(  \frac{dx}{ds}\right)
\nonumber\\
& \nonumber\\
&  =\frac{ds}{dt}\frac{d}{dt}\left(  \frac{ds}{dt}\right)  \frac{dt}{ds}%
\frac{dx}{ds}+\frac{d^{2}x}{ds^{2}}\left(  \frac{ds}{dt}\right)  ^{2}\text{,}%
\end{align}
that is,%
\begin{equation}
\frac{d^{2}x}{dt^{2}}=\frac{d^{2}s}{dt^{2}}\frac{dx}{ds}+\left(  \frac{ds}%
{dt}\right)  ^{2}\frac{d^{2}x}{ds^{2}}\text{.} \label{secondd}%
\end{equation}
Substituting Eqs. (\ref{firstd}) and (\ref{secondd}) into Eq. (\ref{original}%
), we obtain%
\begin{equation}
\left(  \frac{ds}{dt}\right)  ^{2}\frac{d^{2}x}{ds^{2}}+\frac{d^{2}s}{dt^{2}%
}\frac{dx}{ds}+p\left(  t\right)  \frac{ds}{dt}\frac{dx}{ds}+q\left(
t\right)  x=0\text{,}%
\end{equation}
that is, after some algebraic manipulations,%
\begin{equation}
\frac{d^{2}x}{ds^{2}}+\frac{\frac{d^{2}s}{dt^{2}}+p\left(  t\right)  \frac
{ds}{dt}}{\left(  \frac{ds}{dt}\right)  ^{2}}\frac{dx}{ds}+\frac{q\left(
t\right)  }{\left(  \frac{ds}{dt}\right)  ^{2}}x=0\text{,}%
\end{equation}
where $x=x\left(  s\right)  $. Let us define the quantities $A$ and $B$ as,%
\begin{equation}
A\overset{\text{def}}{=}\frac{q\left(  t\right)  }{\left(  \frac{ds}%
{dt}\right)  ^{2}}\text{, and }B\overset{\text{def}}{=}\frac{\frac{d^{2}%
s}{dt^{2}}+p\left(  t\right)  \frac{ds}{dt}}{\left(  \frac{ds}{dt}\right)
^{2}}\text{,} \label{ab}%
\end{equation}
respectively. If we are able to select a suitable change of independent
variables $t\rightarrow s\overset{\text{def}}{=}f\left(  t\right)  $ such that
both $A$ and $B$ are constant quantities, integration of Eq. (\ref{original})
reduces to integration of the following second-order linear differential
equation with constant coefficients:%
\begin{equation}
\frac{d^{2}x}{ds^{2}}+B\frac{dx}{ds}+Ax=0\text{.} \label{pappaciccia}%
\end{equation}
We recall that for $B^{2}<4A$, the system that evolves according to Eq.
(\ref{pappaciccia}) exhibits an under-damped oscillatory motion. Instead, when
$B^{2}>4A$, the system manifests over-damped motion. Finally, when $B^{2}=4A$,
the system is characterized by a critically damped motion.\begin{figure}[ptb]
\centering
\includegraphics[width=0.35\textwidth] {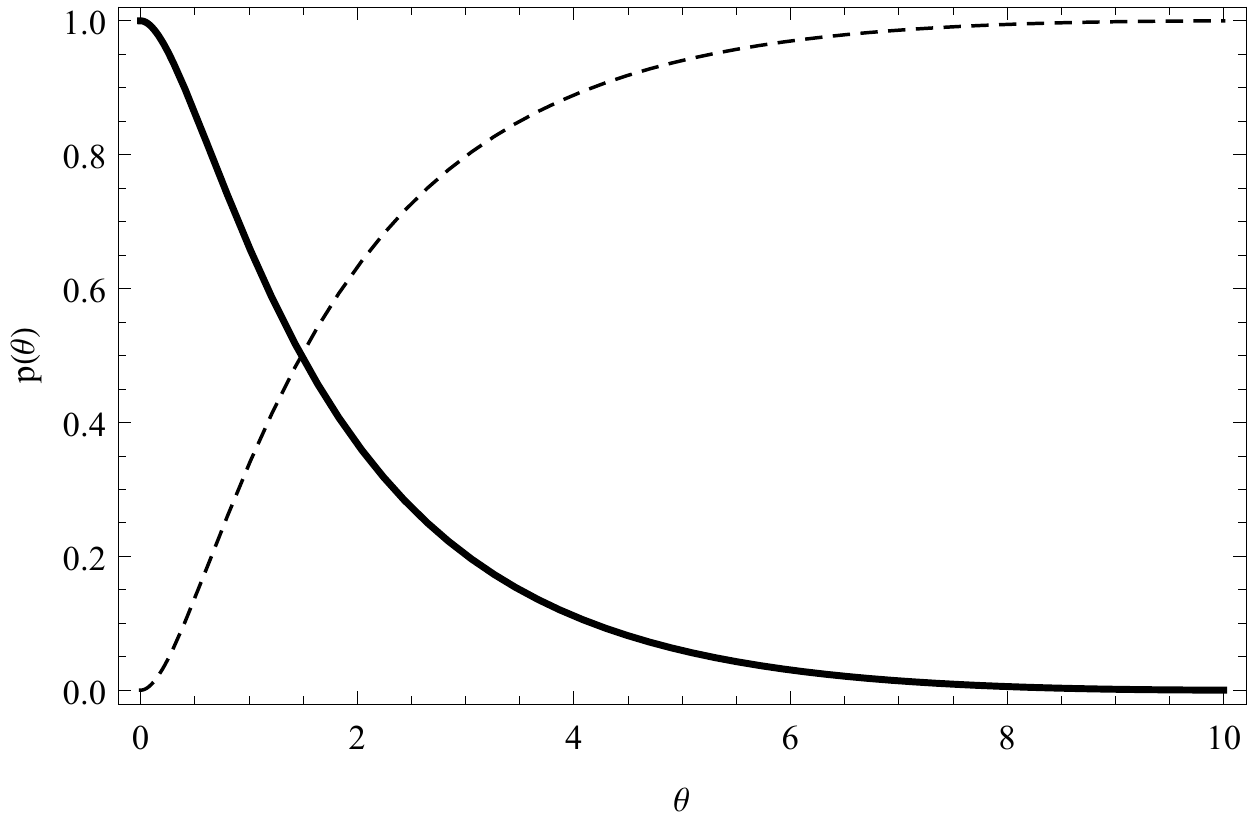}\caption{Monotonic behavior of
the success (dotted) and failure (solid) probabilities in the case of
power-law decay of the Fisher information.}%
\label{fig3}%
\end{figure}We can now return to our problem of integrating Eq. (\ref{powere})
and exploit the above mentioned mathematical reasoning. From Eqs.
(\ref{fsode}) and (\ref{original}), replacing the independent variable $t$
with $\theta$, we have
\begin{equation}
p\left(  \theta\right)  \overset{\text{def}}{=}-\frac{1}{2}\frac
{\mathcal{\dot{F}}}{\mathcal{F}}\text{, and }q\left(  \theta\right)
\overset{\text{def}}{=}\lambda_{\text{FS}}\mathcal{F}^{\frac{1}{2}}\text{.}
\label{pqe}%
\end{equation}
Substituting Eq. (\ref{pqe}) into Eq. (\ref{ab}) and imposing that $A$ and $B$
are constant coefficients, we obtain, after some algebra, the following
suitable change of independent variables,%
\begin{equation}
\theta\rightarrow s:s\left(  \theta\right)  \overset{\text{def}}{=}\frac{1}%
{B}\log\left[  1+\frac{B}{\sqrt{A}}\sqrt{\lambda_{\text{FS}}}\mathcal{F}%
_{0}^{\frac{1}{4}}\theta\right]  \text{,} \label{change}%
\end{equation}
together with the following two-parameter functional form for the Fisher
information function $\mathcal{F}\left(  \theta\text{; }A\text{, }B\right)  $,%
\begin{equation}
\mathcal{F}\left(  \theta\text{; }A\text{, }B\right)  \overset{\text{def}}%
{=}\frac{\mathcal{F}_{0}}{\left[  1+\frac{B}{\sqrt{A}}\sqrt{\lambda
_{\text{FS}}}\mathcal{F}_{0}^{\frac{1}{4}}\theta\right]  ^{4}}\text{.}%
\end{equation}
In summary, we have shown that if $n=4$ and $\Omega$ in Eq. (\ref{power}) is
defined as,%
\begin{equation}
\Omega=\Omega\left(  A\text{, }B\right)  \overset{\text{def}}{=}\frac{B}%
{\sqrt{A}}\sqrt{\lambda_{\text{FS}}}\mathcal{F}_{0}^{\frac{1}{4}}\text{,}%
\end{equation}
there is a suitable change of independent variables defined in Eq.
(\ref{change}) that makes Eq. (\ref{powere}) a linear second-order
differential equation with constant coefficients. Thus, it can now be
integrated in a straightforward manner. For instance, in the case of critical
damping where $B^{2}=4A$ in Eq. (\ref{pappaciccia}), the general solution for
the geodesic path of quantum-mechanical probability amplitudes $q_{\bar{k}%
}\left(  \theta\right)  $ becomes,%
\begin{equation}
q_{\bar{k}}\left(  \theta\right)  \overset{\text{def}}{=}\frac{c_{\bar{k}%
}^{\left(  1\right)  }+c_{\bar{k}}^{\left(  2\right)  }\frac{1}{B}\log\left(
1+\frac{B}{\sqrt{A}}\sqrt{\lambda_{\text{FS}}}\mathcal{F}_{0}^{\frac{1}{4}%
}\theta\right)  }{\left[  1+\frac{B}{\sqrt{A}}\sqrt{\lambda_{\text{FS}}%
}\mathcal{F}_{0}^{\frac{1}{4}}\theta\right]  ^{\frac{1}{2}}}\text{,}%
\end{equation}
where $c_{\bar{k}}^{\left(  1\right)  }$ and $c_{\bar{k}}^{\left(  2\right)
}$ are two real integration constants. In Fig. $3$, setting $A=\frac{1}{4}$,
$B=1$, $\mathcal{F}_{0}=1$, and preserving the normalization constraint, we
observe a monotonic behavior of the success (dotted) and failure (solid)
probabilities in the case of power-law decay of the Fisher information.

\section{On physical systems, Fisher information, and geodesic paths}

In this section, we present some clarifying remarks on the link among physical
systems, Fisher information functions, and geodesic paths on Riemannian manifolds.

\subsection{General remarks}

We point out that classical Fisher information can be computed by considering
parametric probability distributions $p_{\theta}\left(  x\right)
\overset{\text{def}}{=}\left\vert \psi_{\theta}\left(  x\right)  \right\vert
^{2}$ that emerge from the absolute square of parametric quantum mechanical
wavefunctions $\psi_{\theta}\left(  x\right)  $. Similarly, quantum Fisher
information can be defined by means of parametric rank-one projections that
can be regarded as density operators $\rho_{\theta}\left(  x\right)  $
constructed from the above mentioned wavefunctions $\psi_{\theta}\left(
x\right)  $. The functional form of such parametric quantum mechanical
wavefunctions $\psi_{\theta}\left(  x\right)  $ depends on the particular
choice of the unitary evolution operator $U_{\theta}\left(  t\right)  $. The
operator $U_{\theta}\left(  t\right)  $ is generated by the
parameter-dependent Hamiltonian $\mathcal{H}_{\theta}\left(  t\right)  $ that
specifies the physical system under consideration. Furthermore, $\mathcal{H}%
_{\theta}\left(  t\right)  $ acts as the Hermitian generator of temporal
displacements, and satisfies the relation $\mathcal{H}_{\theta}\left(
t\right)  U_{\theta}\left(  t\right)  =i\hbar\partial_{t}U_{\theta}\left(
t\right)  $ with $\partial_{t}\overset{\text{def}}{=}\partial/\partial_{t}$.
The value of the parameter of interest\textbf{ }$\theta$\textbf{ }that
specifies the Hamiltonian\textbf{ }$\mathcal{H}_{\theta}\left(  t\right)  $ is
inferred by observing the evolution of the probe system due to $\mathcal{H}%
_{\theta}\left(  t\right)  $. More specifically, the observation of the probe
system requires finding measurements that are capable of optimally resolving
parameter-dependent neighboring quantum states. Such an optimal resolution is
achieved by employing statistical distinguishability in order to define a
Riemannian metric on the manifold of quantum mechanical density operators.
Then, the Fisher information appears in the infinitesimal line element
$ds_{\text{FS}}^{2}$ on such a manifold, namely $ds_{\text{FS}}^{2}%
\overset{\text{def}}{=}\left(  1/4\right)  \left\{  \mathcal{F}\left(
\theta\right)  +4\sigma_{\dot{\phi}}^{2}\right\}  d\theta^{2}$. Finally, by
integrating the geodesic equations on this Riemannian manifold, one can obtain
the geodesic paths for the probability amplitude variables $q_{\theta}\left(
x\right)  $\textbf{ }with $p_{\theta}\left(  x\right)  \overset{\text{def}}%
{=}q_{\theta}^{2}\left(  x\right)  $.

For pure states, the Fisher information $\mathcal{F}_{\theta}\left(  t\right)
$\textbf{ }reduces to a multiple of the variance $\sigma_{T_{\theta}}^{2}$ of
the Hermitian generator $T_{\theta}$ of displacements in $\theta$.
Specifically,\textbf{ }$\mathcal{F}_{\theta}\left(  t\right)  =4\sigma
_{T_{\theta}}^{2}\overset{\text{def}}{=}4\left\langle \left(  T_{\theta
}-\left\langle T_{\theta}\right\rangle \right)  ^{2}\right\rangle $ with
$T_{\theta}\left(  t\right)  \overset{\text{def}}{=}i\left[  \partial_{\theta
}U_{\theta}\left(  t\right)  \right]  U_{\theta}^{\dagger}\left(  t\right)  $,
$\partial_{\theta}\overset{\text{def}}{=}\partial/\partial_{\theta}$\textbf{,
}and $\left\langle T_{\theta}\right\rangle \overset{\text{def}}{=}%
\mathrm{tr}\left(  \rho_{\theta}T_{\theta}\right)  $.\textbf{ }To simplify the
discussion throughout, we refer to a single parameter of interest $\theta
$\textbf{. }However, our analysis can be generalized in principle to multiple
parameters of interest. In general, the selected parameters of interest are
experimentally controllable quantities. For instance, external magnetic field
intensity, phase difference, temperature, spin-spin coupling constant, volume
per particle, reciprocal temperature, and computing time are all suitable
examples of experimentally controllable parameters of interest. In particular,
for probe systems such as Bose-Einstein condensates and nanomagnetic bits, the
external magnetic field is usually used as a parameter of interest. For a
quantum oscillator in the presence of dephasing noise, the phase difference
plays the role of the parameter of interest. Moreover, temperature, spin-spin
coupling constant, and external magnetic field intensity are three convenient
parameters of interest for Ising spin models. For both a classical ideal gas
and a van der Waals gas, the volume per particle and the reciprocal
temperature $\beta\overset{\text{def}}{=}\left(  k_{B}T\right)  ^{-1}$ with
$k_{B}$ denoting the Boltzmann constant are convenient control parameters.
Finally, for probe systems described by quantum search Hamiltonians, the
computing time can play the role of the parameter of interest.

\subsection{From thermodynamics to quantum metrology}

The Fisher information can assume a variety of functional forms with respect
to the parameter of interest. For instance, within the Fisher information
approach to thermodynamics via the Schr\"{o}dinger equation
\cite{frieden2002,flego03,pennini04}, it can be shown that the Fisher
information $\mathcal{F}_{\text{HO}}\left(  \theta\right)  $ that emerges from
the thermal description of the one-dimensional quantum mechanical harmonic
oscillator is proportional to the harmonic oscillator's specific heat\textbf{
}$C_{V}$\textbf{ }\cite{pennini04}\textbf{,}%
\begin{equation}
\mathcal{F}_{\text{HO}}\left(  \theta\right)  =C_{V}\frac{e^{-\hbar
\omega\theta}}{\theta^{2}}\text{.} \label{fho2}%
\end{equation}
In Eq. (\ref{fho2}), the parameter\textbf{ }$\theta$ denotes the reciprocal
temperature $\beta$ while $\omega$ is the frequency of the oscillator.

In the framework of quantum metrology for a general Hamiltonian parameter
$\mathcal{H}_{\theta}$ \cite{giovannetti06,pang14}, it happens that the
maximum quantum Fisher information is given by%
\begin{equation}
\mathcal{F}_{\max}\left(  \theta\right)  \overset{\text{def}}{=}\left[
\lambda_{\max}\left(  h_{\theta}\right)  -\lambda_{\min}\left(  h_{\theta
}\right)  \right]  ^{2}\text{.} \label{fetamax}%
\end{equation}
The quantities $\lambda_{\max}\left(  h_{\theta}\right)  $ and $\lambda_{\min
}\left(  h_{\theta}\right)  $\textbf{ }denote the maximal and the minimal
eigenvalues of the generator $h_{\theta}$ of parameter translation with
respect to $\theta$,
\begin{equation}
h_{\theta}\overset{\text{def}}{=}i\left(  \partial_{\theta}U_{\theta}\right)
U_{\theta}^{\dagger}\text{,} \label{hteta}%
\end{equation}
with $U_{\theta}\overset{\text{def}}{=}e^{-i\mathcal{H}_{\theta}t}$\textbf{
}in Eq. (\ref{hteta}). For instance, for a spin\textbf{-}$1/2$ particle in an
external magnetic field $\vec{B}\overset{\text{def}}{=}B\hat{n}_{\theta}$ with
$\hat{n}_{\theta}\overset{\text{def}}{=}\left(  \cos\left(  \theta\right)
\text{, }0\text{, }\sin\left(  \theta\right)  \right)  $, the interaction
Hamiltonian $\mathcal{H}_{\theta}$ can be written as%
\begin{equation}
\mathcal{H}_{\theta}\overset{\text{def}}{=}B\left[  \cos\left(  \theta\right)
\sigma_{x}+\sin\left(  \theta\right)  \sigma_{z}\right]  \text{,}
\label{tetah}%
\end{equation}
The Hamiltonian in Eq. (\ref{tetah}) was written by setting the electric
charge $e$, the mass $m$, and the speed of light $c$ equal to one.
Furthermore, $\sigma_{x}$ and $\sigma_{z}$ are Pauli operators. The parameter
$\theta$ is the angle between the $z$ axis and the magnetic field $\vec{B}$.
In this case, it can be shown that $\mathcal{F}_{\max}\left(  \theta\right)  $
in Eq. (\ref{fetamax}) is constant in $\theta$ and equals%
\begin{equation}
\mathcal{F}_{\max}=4B^{2}\sin^{2}\left(  Bt\right)  \text{.}%
\end{equation}
Therefore,\textbf{ }$\mathcal{F}_{\max}$ oscillates with respect to time $t$
and exhibits a period $T\overset{\text{def}}{=}\pi/B$.

For the sake of completeness, we point out that the definition of the Fisher
information is not limited to Hamiltonian systems. For instance, in the
context of an information geometric approach to complex systems in the
presence of limited information \cite{kim11,kim12}, the Fisher information of
Gaussian statistical models is such that\textbf{ }$\mathcal{F}%
_{\text{Gaussian}}\left(  \theta\right)  \propto1/\theta^{2}$ with\textbf{
}$\theta$\textbf{ }denoting the standard deviation of the zero mean
one-dimensional Gaussian random variable that specifies the statistical model
being considered.

\subsection{Analog quantum search}

The information geometric analysis performed in this paper can be especially
relevant to the quantum search problem \cite{grover97}. We recall that
Grover's original quantum search algorithm can be viewed as a definite
discrete-time sequence of elementary unitary transformations acting on qubits
from a digital quantum computing perspective. In particular, given an initial
input state, the output of the algorithm becomes the input state after the
action of the sequence of unitary transformations. Furthermore, the length of
the algorithm is equal to the number of unitary transformations that
characterize the quantum computational software. Finally, the failure
probability after $k$-iterations of Grover's original search algorithm
periodically oscillates as $k$ increases. In Ref. \cite{farhi98}, an analog
counterpart of Grover's algorithm was proposed. The search problem was recast
in terms of finding the normalized target eigenstate $\left\vert
w\right\rangle $ corresponding to the only nonzero eigenvalue $E$ of an
Hamiltonian $\mathcal{H}_{w}\overset{\text{def}}{=}E\left\vert w\right\rangle
\left\langle w\right\vert $ acting on a \emph{complex} $N$-dimensional Hilbert
space. The search ends when the system evolves from the initial state
$\left\vert s\right\rangle $ into the state $\left\vert w\right\rangle $ with
quantum overlap $x=\cos\left(  \theta\right)  \overset{\text{def}}%
{=}\left\langle s|w\right\rangle $. Such evolution is the continuous-time
quantum mechanical Schr\"{o}dinger evolution under the time-independent
Hamiltonian \cite{farhi98}:%
\begin{equation}
\mathcal{H}_{\text{Farhi-Gutmann}}\overset{\text{def}}{=}E\left\vert
w\right\rangle \left\langle w\right\vert +E\left\vert s\right\rangle
\left\langle s\right\vert \text{.} \label{FG}%
\end{equation}
From a physics standpoint, the Hamiltonian formulation of Grover's search
Hamiltonian can be understood in terms of Rabi oscillations between the source
and the target states \cite{byrnes18}. We emphasize that it is possible to
consider a generalized version of $\mathcal{H}_{\text{Farhi-Gutmann}}$ in Eq.
(\ref{FG}) in terms of a more general time-independent quantum search
Hamiltonian\textbf{ }$\mathcal{H}_{\text{oscillation}}$ that describes an
oscillation between the two states $\left\vert s\right\rangle $ and
$\left\vert w\right\rangle $ \cite{bae02},%
\begin{equation}
\mathcal{H}_{\text{oscillation}}\overset{\text{def}}{=}E\left[  \alpha
\left\vert w\right\rangle \left\langle w\right\vert +\beta\left\vert
w\right\rangle \left\langle s\right\vert +\gamma\left\vert s\right\rangle
\left\langle w\right\vert +\delta\left\vert s\right\rangle \left\langle
s\right\vert \right]  \text{,} \label{FGgeneral}%
\end{equation}
where $\alpha$, $\beta$, $\gamma$, $\delta$ are\textbf{ }\emph{complex}
expansion coefficients. We also remark that once the digital-to-analog
transition is completed, information geometry can be employed to view Grover's
iterative procedure as a geodesic path on the manifold of parametric density
operators of pure quantum states built from the continuous approximation of
the parametric quantum output state in Grover's algorithm. In particular, the
Fisher information is computed from the probability distribution vector with
oscillating components that characterize the Groverian geodesic paths and
happens to be constant.

An alternative to Grover's original quantum search algorithm is Grover's
fixed-point search algorithm \cite{grover05}. In particular, the failure
probability after $k$-recursive steps of such algorithm decreases
monotonically and converges to zero as $k$ increases. An analog counterpart of
a fixed-point search algorithm can be recovered by considering time-dependent
Hamiltonians for both fixed-point nonadiabatic \cite{perez07} and adiabatic
\cite{dalzell17} quantum search. These time-dependent Hamiltonians can be
recast as,%
\begin{equation}
\mathcal{H}_{\text{fixed-point}}\left(  t\right)  \overset{\text{def}}{=}%
f_{1}\left(  t\right)  \left[  I-\left\vert s\right\rangle \left\langle
s\right\vert \right]  +f_{2}\left(  t\right)  \left[  I-\left\vert
w\right\rangle \left\langle w\right\vert \right]  \text{,} \label{fixed-point}%
\end{equation}
where $\left\vert s\right\rangle $ is the initial state of the quantum system,
$\left\vert w\right\rangle $ is the target state and $I$ denotes the identity
operator. In the framework of adiabatic quantum search, $f_{1}\left(
t\right)  \overset{\text{def}}{=}1-s\left(  t\right)  $ and $f_{2}\left(
t\right)  \overset{\text{def}}{=}s\left(  t\right)  $ with $s\left(  t\right)
$ being the so-called schedule of the search algorithm. It was shown in Ref.
\cite{dalzell17} that for a suitable choice of parameters that parametrize the
schedule $s\left(  t\right)  $, the Hamiltonian\textbf{ }$\mathcal{H}\left(
t\right)  $ can exhibit both a Grover-like scaling and the fixed-point
property. In particular, such Hamiltonian can drive the system toward a fixed
point. Once the digital-to-analog transition is performed, information
geometry could be exploited to regard Grover's fixed-point algorithm recursive
procedure as a geodesic path on the manifold of parametric density operators
of pure quantum states built from the continuous approximation of the
parametric quantum output state in Grover's fixed point algorithm. In
particular, the Fisher information is computed from the probability
distribution vector with non-oscillating components that characterize the
fixed-point Groverian geodesic paths and happens to be monotonically
decreasing with respect to the parameter of interest chosen to parametrize the
geodesic paths on the underlying manifold. For a recent preliminary
investigation of these ideas, we refer to Ref. \cite{cafaro17}.

In view of these considerations, we have considered in this paper functional
forms of the Fisher information that could be of relevance in the framework of
analog quantum search with search Hamiltonians given in Eqs. (\ref{FG}),
(\ref{FGgeneral}), and (\ref{fixed-point}). More specifically, the quantum
mechanical evolution under the Grover-like search Hamiltonians (GSH) in Eqs.
(\ref{FG}) and (\ref{FGgeneral}) generate wavefunctions that lead to
periodically oscillating probability distributions with constant Fisher
information. Instead, the quantum mechanical evolution under the fixed-point
search Hamiltonian (FPSH) in Eqs. (\ref{fixed-point}) can generate
wavefunctions that lead to monotonically convergent probability distributions
with decreasing Fisher information. Clearly, a deeper understanding of the
connection between the Fisher information and the schedule of the quantum
algorithm remains to be uncovered in order to provide a rigorous mapping
between our geometric analysis and the Hamiltonian formulation of the problem.
In particular, it remains to be understood how to exactly quantify the speed
at which the Hamiltonian can drive the system toward the target state (that
is, the soft or strong nature of monotonic convergence toward the target
state) is related to both the functional forms of the schedule and the Fisher information.

Despite these unresolved issues, we believe that our work represents a
nontrivial step forward towards the accomplishment of such challenging goals.
We also remark that our information geometric analysis can be extended in a
number of ways. For instance, we limited our analysis to a single parameter of
interest and, in addition, we considered only special monotonically decreasing
Fisher information functions. However, the extension of our work to arbitrary
functional forms of the Fisher information, depending or not on more than one
parameter of interest, seems to be outside the reach of analytical treatment.
In this regard, it may be helpful to familiarize with recent numerical
strategies to find optimal protocols as geodesics on a Riemannian manifold
\cite{eric17}. In particular, an oscillating Fisher information\textbf{
}$\mathcal{F}\left(  \theta\right)  $ would require the integration of a
differential equation that describes a damped harmonic oscillator with
$\theta$-dependent damping coefficient given in terms of $\mathcal{\dot{F}%
}/\mathcal{F}$ with $\mathcal{\dot{F}}\overset{\text{def}}{=}d\mathcal{F}%
/d\theta$. In this respect, it may be useful to better understand the very
recent asymptotic stability property for such a type of differential equation
\cite{hatvani18}.

For the time being, we remark that in the case of constant Fisher information,
one deals with geodesic paths that satisfy a differential equation that
formally resembles that of a simple harmonic oscillator and obtains
oscillatory output probabilities. In the case of exponential decay of the
Fisher information, one observes geodesic paths that satisfy a differential
equation that formally resembles that of an aging spring in the presence of
damping together with monotonic output probabilities. Finally, when the Fisher
information exhibits a power-law decay, geodesic paths satisfying an ordinary
differential equation that resembles that of a critically damped harmonic
oscillator and leads to monotonic output probabilities. The presence of
damping effects seems to lead to the characteristic monotonic behavior of the
quantum mechanical probability amplitude squared. Therefore, it is reasonable
to further investigate this plausible connection between Fisher information
and dissipative effects in an effort to render any such connection more
rigorous. This latter point shall be addressed in the next section by
exploiting a Riemannian geometric characterization of thermodynamic concepts.

\section{Riemannian geometric viewpoint of thermodynamic concepts}

An efficient thermodynamic process occurs by minimum dissipation or maximum
power. In particular, dissipation can be quantified in terms of the amount of
work lost in the process \cite{salamon83}. Availability loss (that is,
dissipated availability or irreversibility) and entropy production are the two
most common measures of dissipation in thermodynamics \cite{hoffmann89}. In a
Riemannian approach to thermodynamics, both availability loss and entropy
production are related to the concept of thermodynamic length. However, while
in the former case one deals with the so-called energy version of the
thermodynamic length, in the latter case the so-called entropy version of the
thermodynamic length is taken into consideration \cite{salamon84}.
Specifically, optimum paths that minimize entropy production are commonly
referred to as \emph{optimum cooling paths} (that is, maximum reversibility
paths) and characterize a thermodynamic process that occurs at constant
thermodynamic speed \cite{andresen94,spirkl95,diosi96,diosi2000}. The notions
of thermodynamic length and dissipated availability will be discussed in the
next subsection.

\subsection{Preliminaries}

Consider a physical system, small in mass and extent, surrounded by an
(infinite) environment with temperature $T_{0}$ and pressure $p_{0}$ which are
unaffected by any process experienced by the system. An arbitrary process can
be viewed as an interaction between the system and the environment, once one
includes in the system as much material or machinery that is affected by the
process itself. Under these working assumptions, Gibbs introduced a quantity
$\Phi$ (that is, the Gibbs free energy \cite{huang87}) defined as,%
\begin{equation}
\Phi\overset{\text{def}}{=}E+p_{0}V-T_{0}S\text{,}%
\end{equation}
where $E$ is the energy of the system, $V$ is its volume, $S$ denotes its
entropy, and Gibbs showed that (for further details, see Ref.\cite{gibbs28})
\begin{equation}
\Delta\Phi\leq0\text{,}%
\end{equation}
where $\Delta\Phi$ is the increase in the quantity $\Phi$. The availability
(or, available energy) $\Lambda$ of the system and the environment is defined
as \cite{keenan51},%
\begin{equation}
\Lambda\overset{\text{def}}{=}\Phi-\Phi_{\min}\text{,} \label{availability}%
\end{equation}
where $\Phi_{\min}$ is the minimum possible value of $\Phi$ attained when the
system is in a state from which no spontaneous changes can happen. Such a
state of the system is the state of stable equilibrium (or, more generally,
maximum stability) and is characterized by a pressure $p_{0}$ and a
temperature $T_{0}$. The availability $\Lambda$ in Eq. (\ref{availability})
represents the maximum value of the useful work, that is to say, work in
excess of that done against the environment that could be obtained from the
system and the environment via any arbitrary process, without intervention of
other bodies:%
\begin{equation}
\Lambda=W_{\text{excess}}\text{.}%
\end{equation}
We point out that for the most stable state of the system, $\Lambda=0$.
Furthermore, for any state of any system immersed in a stable environment,
$\Lambda\geq0$.

Transitioning from a conventional to a geometrical setting, the so-called
thermodynamic length $\mathcal{L}_{\text{th.}}$ of a curve $\theta^{\mu
}=\theta^{\mu}\left(  t\right)  $ parametrized by $t$ with $0\leq t\leq\tau$
is defined as \cite{salamon83},%
\begin{equation}
\mathcal{L}_{\text{th.}}\overset{\text{def}}{=}\int_{0}^{\tau}\sqrt{g_{\mu\nu
}\left(  \theta\right)  \frac{d\theta^{\mu}}{dt}\frac{d\theta^{\nu}}{dt}%
}dt\text{,} \label{length}%
\end{equation}
where $g_{\mu\nu}\left(  \theta\right)  $ in Eq. (\ref{length}) denotes the
so-called thermodynamic metric tensor given by \cite{schlogl85, brody95},%
\begin{equation}
g_{\mu\nu}\left(  \theta\right)  \overset{\text{def}}{=}\frac{\partial^{2}%
\psi}{\partial\theta^{\mu}\partial\theta^{\nu}}=-\frac{\partial\left\langle
X_{\mu}\right\rangle }{\partial\theta^{\nu}}=\left\langle \left(  X_{\mu
}-\left\langle X_{\mu}\right\rangle \right)  \left(  X_{\nu}-\left\langle
X_{\nu}\right\rangle \right)  \right\rangle \text{.} \label{tmt}%
\end{equation}
The quantity $\psi$ in Eq. (\ref{tmt}) denotes the free entropy,%
\begin{equation}
\psi\overset{\text{def}}{=}\log\left(  \mathcal{Z}\right)  =-\beta
\Phi=\mathcal{S-}\theta^{\mu}\left\langle X_{\mu}\right\rangle \text{,}
\label{free}%
\end{equation}
where $\mathcal{Z}$, $\Phi$, $\mathcal{S}$, $\beta\overset{\text{def}}{=}%
\frac{1}{k_{\text{B}}T}$, and $k_{\text{B}}$ are the partition function, free
energy, entropy, reciprocal temperature ($T$), and Boltzmann constant,
respectively. The variables $\left\{  X_{\mu}\left(  x\right)  \right\}  $ are
thermodynamic variables that specify the Hamiltonian of the system (for
instance, internal energy and volume) while $x$ belongs to the configuration
space. Furthermore, the time-dependent $\theta$'s are experimentally
controllable parameters of the system that specify the accessible
thermodynamic state space of the system. Expectation values in Eqs.
(\ref{tmt}) and (\ref{free}) are defined with respect to the probability
distribution $p\left(  x|\theta\right)  $ (Gibbs ensemble),%
\begin{equation}
p\left(  x|\theta\right)  \overset{\text{def}}{=}\frac{1}{\mathcal{Z}%
}e^{-\beta\mathcal{H}\left(  x\text{, }\theta\right)  }=\frac{1}{\mathcal{Z}%
}e^{-\theta^{\mu}\left(  t\right)  X_{\mu}\left(  x\right)  }\text{,}
\label{Gprobability}%
\end{equation}
where, adopting the Einstein convention, repeated lower and upper indices are
summed over. We point out that, using Eqs. (\ref{tmt}), (\ref{free}), and
(\ref{Gprobability}), the thermodynamic metric tensor can be shown to be equal
to the Fisher-Rao information metric tensor:%
\begin{equation}
g_{\mu\nu}\left(  \theta\right)  \overset{\text{def}}{=}\frac{\partial^{2}%
\psi}{\partial\theta^{\mu}\partial\theta^{\nu}}=\int p\left(  x|\theta\right)
\frac{\partial\log p\left(  x|\theta\right)  }{\partial\theta^{\mu}}%
\frac{\partial\log p\left(  x|\theta\right)  }{\partial\theta^{\nu}}dx\text{.}
\label{thermofisher}%
\end{equation}
The quantity $g_{\mu\nu}\left(  \theta\right)  $ in Eq. (\ref{thermofisher})
is a Riemannian metric on the manifold of thermodynamic states. The
thermodynamic length in Eq. (\ref{length}) has dimensions of speed and its
physical interpretation is related to the concept of availability loss (or,
dissipated availability) $\Lambda_{\text{dissipated}}$ in a thermodynamic
process \cite{salamon83,hoffmann89}:%
\begin{equation}
\Lambda_{\text{dissipated}}\overset{\text{def}}{=}\int_{0}^{\tau}g_{\mu\nu
}\left(  \theta\right)  \frac{d\theta^{\mu}}{dt}\frac{d\theta^{\nu}}%
{dt}dt\text{.} \label{da}%
\end{equation}
The quantity $\Lambda_{\text{dissipated}}$ can be expressed in terms of the
so-called thermodynamic divergence of the path $\mathcal{D}$
\cite{crooks07,sivak12},%
\begin{equation}
\mathcal{D}\overset{\text{def}}{=}\tau\cdot\Lambda_{\text{dissipated}}\text{,}%
\end{equation}
where, in the context of Riemannian geometry, $\mathcal{D}/2\tau$ is also
known as the energy of the path parametrized with $t$ where $0\leq t\leq\tau$.
Indeed, considering Eqs. (\ref{length}) and (\ref{da}), the application of the
Cauchy-Schwarz inequality leads to the following inequality%
\begin{equation}
\Lambda_{\text{dissipated}}\geq\frac{\mathcal{L}_{\text{th.}}^{2}}{\tau
}\text{,} \label{inequality}%
\end{equation}
that is, $\mathcal{D}\geq\mathcal{L}_{\text{th.}}^{2}$ (the divergence-length
inequality expresses the fact that the minimum divergence of the path is the
square of the thermodynamic length). The equality in Eq. (\ref{inequality}) is
obtained only for the most favorable time parametrization, which occurs when%
\begin{equation}
\left\Vert \dot{\theta}\left(  t\right)  \right\Vert \overset{\text{def}}%
{=}\left(  g_{\mu\nu}\left(  \theta\right)  \frac{d\theta^{\mu}}{dt}%
\frac{d\theta^{\nu}}{dt}\right)  ^{\frac{1}{2}}=\text{const,}%
\end{equation}
with the constant equal to $\mathcal{L}_{\text{th.}}/\tau$.\ Therefore, a
thermodynamic process dissipates minimum availability when it proceeds at
constant speed.

\subsection{Illustrative examples}

Given the equivalence between the Fisher-Rao information metric and the
thermodynamic metric tensor, we can apply the concepts of thermodynamic length
and availability loss to our selected illustrative examples discussed in Sec.
V. We recall that our output probability paths $p_{\bar{k}}\left(
\theta\right)  $ are parametrized by a single statistical parameter $\theta$
that denotes the computational time of a quantum process.

In general, the geodesic equations satisfied by the statistical parameters
$\theta^{\mu}=\theta^{\mu}\left(  t\right)  $ with $1\leq\mu\leq\left\vert
\Theta\right\vert $, where $\left\vert \Theta\right\vert $ denotes the
cardinality of the set $\Theta$ of statistical parameters, are given by,%
\begin{equation}
\frac{d^{2}\theta^{\mu}}{dt^{2}}+\Gamma_{\nu\rho}^{\mu}\frac{d\theta^{\nu}%
}{dt}\frac{d\theta^{\rho}}{dt}=0\text{.} \label{otto3}%
\end{equation}
The quantities $\Gamma_{\nu\rho}^{\mu}$ in Eq. (\ref{otto3}) are the
connection coefficients defined as,%
\begin{equation}
\Gamma_{\nu\rho}^{\mu}\overset{\text{def}}{=}\frac{1}{2}g^{\mu\alpha}\left(
\partial_{\nu}g_{\alpha\rho}+\partial_{\rho}g_{\nu\alpha}-\partial_{\alpha
}g_{\nu\rho}\right)  \text{,} \label{otto2}%
\end{equation}
where $\partial_{\nu}\overset{\text{def}}{=}\frac{\partial}{\partial
\theta^{\nu}}$. In our analysis, we have%
\begin{equation}
ds_{\text{FS}}^{2}=g_{\theta\theta}\left(  \theta\right)  d\theta^{2}\text{,
with }g_{\theta\theta}\left(  \theta\right)  \overset{\text{def}}{=}\frac
{1}{4}\mathcal{F}\left(  \theta\right)  \text{.} \label{otto1}%
\end{equation}
Using Eqs. (\ref{otto1}) and (\ref{otto2}), the geodesic equation in Eq.
(\ref{otto3}) becomes%
\begin{equation}
\frac{d^{2}\theta}{dt^{2}}+\frac{1}{2\mathcal{F}}\frac{d\mathcal{F}}{d\theta
}\left(  \frac{d\theta}{dt}\right)  ^{2}=0\text{.} \label{odeti}%
\end{equation}
From the integration of Eq. (\ref{odeti}), we can also consider the so-called
computational speed defined as,
\begin{equation}
v\left(  t\right)  \overset{\text{def}}{=}\left(  g_{\theta\theta}\left(
\theta\right)  \left(  \frac{d\theta}{dt}\right)  ^{2}\right)  ^{\frac{1}{2}%
}=\frac{1}{2}\sqrt{\mathcal{F}\left(  \theta\left(  t\right)  \right)  }%
\frac{d\theta}{dt}\text{.} \label{speed}%
\end{equation}
In what follows, we compute the availability loss $\Lambda_{\text{dissipated}%
}$ in Eq. (\ref{da}) and the computational speed $v$ in Eq. (\ref{speed})
after integrating the nonlinear ordinary differential equation in
(\ref{odeti}) whose structure clearly depends on the functional form of the
Fisher information function $\mathcal{F}$. Below, we consider the three cases
considered in Sec. V.

\subsubsection{Example one: Constant Fisher information}

In this case, since $\mathcal{F}\left(  \theta\right)  \overset{\text{def}}%
{=}\mathcal{F}_{0}$, Eq. (\ref{odeti}) becomes%
\begin{equation}
\frac{d^{2}\theta}{dt^{2}}=0\text{.}%
\end{equation}
Assuming as initial conditions $\theta\left(  t_{0}\right)  =\theta_{0}$ and
$\dot{\theta}\left(  t_{0}\right)  =\dot{\theta}_{0}$, we obtain%
\begin{equation}
\theta\left(  t\right)  =\theta_{0}+\dot{\theta}_{0}\left(  t-t_{0}\right)
\text{.}%
\end{equation}
Furthermore, the availability loss $\Lambda_{\text{dissipated}}$ in
Eq.(\ref{da}) becomes%
\begin{equation}
\Lambda_{\text{dissipated}}\left(  \tau\right)  =\frac{\mathcal{F}_{0}}{4}%
\dot{\theta}_{0}^{2}\tau\text{.} \label{lambda1}%
\end{equation}
Finally, the computational speed $v$ in Eq. (\ref{speed}) is given by%
\begin{equation}
v=\frac{1}{2}\sqrt{\mathcal{F}_{0}}\dot{\theta}_{0}\text{.} \label{speed1}%
\end{equation}
We notice that the quantum process proceeds at constant speed and, thus,
dissipates minimum availability. Moreover, the dissipated availability grows
linearly with $\tau$ (that is, the length of the parametrization interval).

\subsubsection{Example two: Exponential decay}

In this case, since $\mathcal{F}\left(  \theta\right)  \overset{\text{def}}%
{=}\mathcal{F}_{0}e^{-\xi\theta}$, Eq. (\ref{odeti}) becomes%
\begin{equation}
\frac{d^{2}\theta}{dt^{2}}-\frac{\xi}{2}\left(  \frac{d\theta}{dt}\right)
^{2}=0\text{.} \label{fuckyou}%
\end{equation}
Assuming initial conditions $\theta\left(  t_{0}\right)  =\theta_{0}$ and
$\dot{\theta}\left(  t_{0}\right)  =\dot{\theta}_{0}$, integration of Eq.
(\ref{fuckyou}) yields%
\begin{equation}
\theta\left(  t\right)  =\theta_{0}-\frac{2}{\xi}\log\left[  1-\xi\frac
{\dot{\theta}_{0}}{2}\left(  t-t_{0}\right)  \right]  \text{.}%
\end{equation}
Furthermore, the availability loss $\Lambda_{\text{dissipated}}$ in Eq.
(\ref{da}) becomes%
\begin{equation}
\Lambda_{\text{dissipated}}\left(  \tau\right)  =\frac{\mathcal{F}_{0}}{4}%
\dot{\theta}_{0}^{2}e^{-\xi\theta_{0}}\tau\text{.} \label{lambda2}%
\end{equation}
Finally, the computational speed $v$ in Eq. (\ref{speed}) is given by%
\begin{equation}
v=\frac{1}{2}\sqrt{\mathcal{F}_{0}}e^{-\frac{\xi}{2}\theta_{0}}\dot{\theta
}_{0}\text{.} \label{speed2}%
\end{equation}
In analogy to the first example, the quantum process proceeds at constant
speed and, thus, dissipates minimum availability. Moreover, the dissipated
availability grows linearly with $\tau$. However, comparing Eqs.
(\ref{lambda1}) and (\ref{speed1}) with Eqs. (\ref{lambda2}) and
(\ref{speed2}), we observe that while the computational speed of the process
is smaller in this second case, the availability loss is also smaller.

\subsubsection{Example three: Power-law decay}

In this case, since $\mathcal{F}\left(  \theta\right)  \overset{\text{def}}%
{=}\frac{\mathcal{F}_{0}}{\left(  1+\Omega\theta\right)  ^{4}}$, Eq.
(\ref{odeti}) becomes%
\begin{equation}
\frac{d^{2}\theta}{dt^{2}}-\frac{2\Omega}{1+\Omega\theta}\left(  \frac
{d\theta}{dt}\right)  ^{2}=0\text{.} \label{domyjob}%
\end{equation}
Assuming initial conditions $\theta\left(  t_{0}\right)  =\theta_{0}$ and
$\dot{\theta}\left(  t_{0}\right)  =\dot{\theta}_{0}$, integrating Eq.
(\ref{domyjob}), we obtain%
\begin{equation}
\theta\left(  t\right)  =\frac{\left(  1+\Omega\theta_{0}\right)  ^{2}%
+\Omega\dot{\theta}_{0}\left[  \left(  t-t_{0}\right)  -\frac{1+\Omega
\theta_{0}}{\Omega\dot{\theta}_{0}}\right]  }{\Omega^{2}\dot{\theta}%
_{0}\left[  \frac{1+\Omega\theta_{0}}{\Omega\dot{\theta}_{0}}-\left(
t-t_{0}\right)  \right]  }\text{.}%
\end{equation}
Furthermore, the availability loss $\Lambda_{\text{dissipated}}$ in Eq.
(\ref{da}) becomes%
\begin{equation}
\Lambda_{\text{dissipated}}\left(  \tau\right)  =\frac{\mathcal{F}_{0}}%
{4}\frac{\dot{\theta}_{0}^{2}}{\left(  1+\Omega\theta_{0}\right)  ^{2}}%
\tau\text{.} \label{lambda3}%
\end{equation}
Finally, the computational speed $v$ in Eq. (\ref{speed}) is given by%
\begin{equation}
v=\frac{1}{2}\sqrt{\mathcal{F}_{0}}\frac{1}{1+\Omega\theta_{0}}\dot{\theta
}_{0}\text{.} \label{speed3}%
\end{equation}
In analogy to the first and second examples, the quantum process proceeds at
constant speed and, thus, dissipates minimum availability. In addition, the
dissipated availability grows linearly with $\tau$. However, comparing Eqs.
(\ref{lambda1}) and (\ref{speed1}) with Eqs. (\ref{lambda3}) and
(\ref{speed3}), we observe that while the computational speed of the process
is smaller in this third case, the availability loss is also smaller. In Table
I, we report the observed behavior of availability losses, computational
speeds, and geodesic paths for different physical scenarios that can arise
from different functional forms of the Fisher information.

\begin{table}[t]
\centering
\begin{tabular}
[c]{c|c|c|c|c|c}\hline\hline
Fisher Information & Geodesic Paths & Physical System & Probability &
Availability Loss & Speed\\\hline
constant & simple harmonic oscillator & GSH & oscillatory & higher & higher\\
exponential decay & aging spring with damping & strong convergence FPSH &
monotonic & lower & lower\\
power law decay & critically damped harmonic oscillator & soft convergence
FPSH & monotonic & lower & lower\\\hline
\end{tabular}
\caption{Behavior of availability losses, computational speeds, and geodesic
paths for different physical scenarios that can arise from different
functional forms of the Fisher information. GSH and FPSH denote Grover-like
search Hamiltonians and fixed-point-like search Hamiltonians, respectively.}%
\end{table}

\section{Concluding Remarks}

In this paper, we presented an information geometric characterization of the
oscillatory or monotonic behavior of statistically parametrized squared
probability amplitudes originating from special functional forms of the Fisher
information function: constant, exponential decay, and power-law decay.
Furthermore, for each case, we computed both the computational speed and the
availability loss of the corresponding physical processes by employing a
convenient Riemannian geometrization of thermodynamical concepts. In what
follows, we outline our main findings in a more detailed fashion:

\begin{enumerate}
\item We provided a dynamical information geometric characterization of the
Fisher information function via an explicit derivation of the Euler-Lagrange
equations satisfied by the quantum-mechanical probability amplitudes of pure
states using variational calculus techniques applied to an action functional
defined in terms of either the Fubini-Study [see Eq. (\ref{fsode})] or the
Wigner-Yanase [see Eq. (\ref{wyode})] metric tensors.

\item We analyzed the parametric behavior of the squared probability
amplitudes arising from three different classes of Fisher information
functions: constant Fisher information, exponential decay, and power-law
decay. In the first case, we observed oscillatory behavior of the output
probabilities (Fig. $1$) that arises from the integration of a differential
equation describing a simple harmonic oscillator [see Eq.(\ref{fsode1})]. In
the second case, we reported monotonic behavior of the output probabilities
(Fig. $2$) that originates from the integration of a differential equation
characterizing an aging spring in the presence of damping [see Eq.
(\ref{fsode2})]. Finally, in the third case, we observed monotonic behavior of
the output probabilities (Fig. $3$). In particular, upon a suitable change of
variables, the reported behavior of the output probabilities can be explained
as emerging from the integration of a differential equation describing a
critically damped harmonic oscillator [see Eqs. (\ref{powere}),
(\ref{pappaciccia}), and (\ref{change})]. The overall picture emerging from
the analysis of these three cases inspired us to further investigate the
connection between the Fisher information and dissipative effects.

\item We used the Riemannian geometrization of thermodynamical concepts,
including thermodynamic speed and dissipated availability, to study the
behavior of both the availability loss [see Eq. (\ref{da})] and the
computational speed [see Eq. (\ref{speed})] of the quantum processes specified
in terms of the previously mentioned output probability paths. Specifically,
after finding the optimal parametrization of the statistical variable $\theta$
that specifies our output probabilities $p_{\bar{k}}\left(  \theta\right)  $,
we evaluated both the availability loss and the computational speed along the
optimal geodesic paths corresponding to the above mentioned three scenarios
[see Table I together with Eqs. (\ref{lambda1}), (\ref{speed1}),
(\ref{lambda2}), (\ref{speed2}), (\ref{lambda3}), and (\ref{speed3})]. Our
main finding here is that a greater computational speed comes necessarily at
the expense of a greater availability loss.
\end{enumerate}

As a final remark, we recall that from a quantum mechanical standpoint, the
output state in Grover's quantum search algorithm follows a geodesic path
obtained from the Fubini-Study metric on the manifold of Hilbert-space rays.
In addition, Grover's algorithm is specified by a constant Fisher information.
A topic of great interest in quantum computing is the investigation of
constructive uses of dissipation. For instance, in Ref. \cite{ari09} it was
shown that it is possible to modify Grover's algorithm by introducing a
suitable amount of dissipation in such a manner that the newly obtained
algorithm, while preserving the typical number of queries $O\left(  \sqrt
{N/M}\right)  $ (where $N$ is the number of items and $M$ is the number of
target items), gains robustness by damping out the oscillations between the
target and nontarget states. Furthermore, the problem of designing quantum
algorithms that are both fast and thermodynamically efficient is a very
challenging and relevant problem \cite{castelvecchi}. To the best of our
knowledge, there does not exist any conclusive investigation that concerns
this type of issue. In Ref. \cite{campbell17}, however, it was shown there
that the faster one seeks to implement a shortcut, the higher is the
thermodynamic cost of realizing the associated quantum process.

Despite the limits of our investigation, we are confident that our information
geometric analysis of the evolution of quantum systems combined with
thermodynamical considerations can be especially relevant to information
physicists and, more specifically, quantum information theorists with
particular interest in thermodynamical aspects of quantum information. We also
strongly believe that the significance of our work runs far deeper than what
is presently understood. However, significant further exploration is needed to
make a precise formal connection among parameter-dependent probe Hamiltonians,
Fisher information, and optimal cooling paths on the underlying parameter
manifold. In conclusion, based also on our recent findings in quantum
computing \cite{cafaro17}, statistical mechanics \cite{cafaropre1,cafaropre2},
and information geometry \cite{cafaroRMP,felice18}, we have reason to believe
that our information geometric analysis presented in this paper will pave the
way to further quantitative investigations on the role played by the Fisher
information function in the trade-off between speed and thermodynamic
efficiency in quantum search algorithms.

\begin{acknowledgments}
C. C. is grateful to the United States Air Force Research Laboratory (AFRL)
Summer Faculty Fellowship Program (SFFP) for providing support for this work.
Any opinions, findings and conclusions or recommendations expressed in this
paper are those of the authors and do not necessarily reflect the views of
AFRL. Finally, constructive criticism from an anonymous referee leading to an
improved version of this paper is sincerely acknowledged by the authors.
\end{acknowledgments}

\pagebreak

\end{document}